\documentclass[twocolumn]{aastex701}

\usepackage{savesym}
\savesymbol{tablenum}
\usepackage{siunitx}
\restoresymbol{SIX}{tablenum}
\usepackage{enumitem}
\usepackage{xspace}

\newcommand{\oii}{\mbox{[\ion{O}{2}]}}
\newcommand{\oiii}{\mbox{[\ion{O}{3}]}}
\newcommand{\hei}{\mbox{\ion{He}{1}}}

\newcommand{\mgii}{\mbox{\ion{Mg}{2}}}

\newcommand{\siii}{\mbox{[\ion{S}{3}]}}

\newcommand{\hb}{\mbox{H$\beta$}}
\newcommand{\ha}{\mbox{H$\alpha$}}
\newcommand{\bra}{\mbox{Br$\alpha$}}

\newcommand{\paa}{\mbox{Pa$\alpha$}}
\newcommand{\pab}{\mbox{Pa$\beta$}}

\newcommand{\lya}{\mbox{Ly$\alpha$}}

\newcommand\msun{\mbox{\si{M_\odot}}}

\newcommand\smpy{\mbox{\si{M_\odot.yr^{-1}}}}

\newcommand\msunsqpc{\mbox{\si{M_\odot.pc^{-2}}}}
\renewcommand{\micron}{\si{\micro\meter}}
\newcommand{\zsp}{\mbox{$z_\mathrm{spec}$}}
\newcommand{\zph}{\mbox{$z_\mathrm{phot}$}}

\usepackage[dvipsnames]{xcolor}

\definecolor{cfsun}{HTML}{be002f}
\newcommand{\fsun}[1]{{#1}}

\begin{document}

\title{JADES: Discovery of Large Reservoirs of Small Dust Grains in the Circumgalactic Medium of Massive Galaxies at $z\sim3.5$ through Deep JWST/NIRCam Imaging and Grism Spectroscopy
}

\author[0000-0002-4622-6617]{Fengwu Sun}
\affiliation{Center for Astrophysics $|$ Harvard \& Smithsonian, 60 Garden St., Cambridge, MA 02138, USA}
\email{fengwu.sun@cfa.harvard.edu}  

\author[0000-0002-2929-3121]{Daniel J.\ Eisenstein}
\affiliation{Center for Astrophysics $|$ Harvard \& Smithsonian, 60 Garden St., Cambridge, MA 02138, USA}
\email{deisenstein@cfa.harvard.edu}  

\author[0000-0003-2388-8172]{Francesco D'Eugenio}
\affiliation{Kavli Institute for Cosmology, University of Cambridge, Madingley Road, Cambridge, CB3 0HA, UK}
\affiliation{Cavendish Laboratory, University of Cambridge, 19 JJ Thomson Avenue, Cambridge, CB3 0HE, UK}
\email{fd391@cam.ac.uk}

\author[0000-0003-4565-8239]{Kevin Hainline}
\affiliation{Steward Observatory, University of Arizona, 933 N. Cherry Avenue, Tucson, AZ 85721, USA}
\email{kevinhainline@arizona.edu}

\author[0000-0003-4337-6211]{Jakob M.\ Helton}
\affiliation{Department of Astronomy \& Astrophysics, The Pennsylvania State University, University Park, PA 16802, USA}
\email{jakobhelton@psu.edu}

\author[0000-0002-9280-7594]{Benjamin D.\ Johnson}
\affiliation{Center for Astrophysics $|$ Harvard \& Smithsonian, 60 Garden St., Cambridge, MA 02138, USA}
\email{benjamin.johnson@cfa.harvard.edu}

\author[0000-0001-6052-4234]{Xiaojing Lin}
\affiliation{Department of Astronomy, Tsinghua University, Beijing 100084, China}
\affiliation{Steward Observatory, University of Arizona, 933 N. Cherry Avenue, Tucson, AZ 85721, USA}
\email{xiaojinglin@arizona.edu}

\author[0000-0002-7893-6170]{Marcia Rieke}
\affiliation{Steward Observatory, University of Arizona, 933 N. Cherry Avenue, Tucson, AZ 85721, USA}
\email{mrieke@gmail.com}

\author[0000-0002-4271-0364]{Brant Robertson}
\affiliation{Department of Astronomy and Astrophysics, University of California, Santa Cruz, 1156 High Street, Santa Cruz, CA 95064, USA}
\email{brant@ucsc.edu}

\author[0000-0002-8224-4505]{Sandro Tacchella}
\affiliation{Kavli Institute for Cosmology, University of Cambridge, Madingley Road, Cambridge, CB3 0HA, UK}
\affiliation{Cavendish Laboratory, University of Cambridge, 19 JJ Thomson Avenue, Cambridge, CB3 0HE, UK}
\email{st578@cam.ac.uk}

\author[0000-0002-8651-9879]{Andrew J.\ Bunker}
\affiliation{Department of Physics, University of Oxford, Denys Wilkinson Building, Keble Road, Oxford OX1 3RH, UK}
\email{andy.bunker@physics.ox.ac.uk}

\author[0000-0002-7636-0534]{Jacopo Chevallard}
\affiliation{Department of Physics, University of Oxford, Denys Wilkinson Building, Keble Road, Oxford OX1 3RH, UK}
\email{chevalla@iap.fr}

\author[0000-0002-9551-0534]{Emma Curtis-Lake}
\affiliation{Centre for Astrophysics Research, Department of Physics, Astronomy and Mathematics, University of Hertfordshire, Hatfield AL10 9AB, UK}
\email{e.curtis-lake@herts.ac.uk}

\author[0000-0003-1344-9475]{Eiichi Egami}
\affiliation{Steward Observatory, University of Arizona, 933 N. Cherry Avenue, Tucson, AZ 85721, USA}
\email{egami@arizona.edu}

\author[0000-0002-8543-761X]{Ryan Hausen}
\affiliation{Department of Physics and Astronomy, The Johns Hopkins University, 3400 N. Charles St., Baltimore, MD 21218}
\email{rhausen@ucsc.edu}

\author[0000-0001-7673-2257]{Zhiyuan Ji}
\affiliation{Steward Observatory, University of Arizona, 933 N. Cherry Avenue, Tucson, AZ 85721, USA}
\email{zhiyuanji@arizona.edu}

\author[0000-0002-6221-1829]{Jianwei Lyu}
\affiliation{Steward Observatory, University of Arizona, 933 N. Cherry Avenue, Tucson, AZ 85721, USA}
\email{jianwei@arizona.edu}

\author[0000-0002-4985-3819]{Roberto Maiolino}
\affiliation{Kavli Institute for Cosmology, University of Cambridge, Madingley Road, Cambridge, CB3 0HA, UK}
\affiliation{Cavendish Laboratory - Astrophysics Group, University of Cambridge, 19 JJ Thomson Avenue, Cambridge, CB3 0HE, UK}
\affiliation{Department of Physics and Astronomy, University College London, Gower Street, London WC1E 6BT, UK}
\email{rm665@cam.ac.uk}

\author[0000-0002-5104-8245]{Pierluigi Rinaldi}
\affiliation{Space Telescope Science Institute, 3700 San Martin Drive, Baltimore, Maryland 21218, USA}
\email{prinaldi@stsci.edu}

\author[0000-0001-6561-9443]{Yang Sun}
\affiliation{Steward Observatory, University of Arizona, 933 N. Cherry Avenue, Tucson, AZ 85721, USA}
\email{sunyang@arizona.edu}

\author[0000-0002-9081-2111]{James A.\ A.\ Trussler}
\affiliation{Center for Astrophysics $|$ Harvard \& Smithsonian, 60 Garden St., Cambridge, MA 02138, USA}
\email{james.trussler@cfa.harvard.edu}

\author[0000-0003-2919-7495]{Christina C.\ Williams}
\affiliation{NSF National Optical-Infrared Astronomy Research Laboratory, 950 North Cherry Avenue, Tucson, AZ 85719, USA}
\email{christina.williams@noirlab.edu}

\author[0000-0001-9262-9997]{Christopher N.\ A.\ Willmer}
\affiliation{Steward Observatory, University of Arizona, 933 N. Cherry Avenue, Tucson, AZ 85721, USA}
\email{cnaw@as.arizona.edu}

\author[0000-0002-7595-121X]{Joris Witstok}
\affiliation{Cosmic Dawn Center (DAWN), Copenhagen, Denmark}
\affiliation{Niels Bohr Institute, University of Copenhagen, Jagtvej 128, DK-2200, Copenhagen, Denmark}
\email{joris.witstok@nbi.ku.dk}

\author[0000-0002-8876-5248]{Zihao Wu}
\affiliation{Center for Astrophysics $|$ Harvard \& Smithsonian, 60 Garden St., Cambridge, MA 02138, USA}
\email{zihao.wu@cfa.harvard.edu}

\author[0000-0003-3307-7525]{Yongda Zhu}
\affiliation{Steward Observatory, University of Arizona, 933 N. Cherry Avenue, Tucson, AZ 85721, USA}
\email{yongdaz@arizona.edu}

\correspondingauthor{Fengwu Sun}
\email[show]{fengwu.sun@cfa.harvard.edu}


\begin{abstract}

Using JWST NIRCam imaging and grism spectroscopy from the JWST Advanced Deep Extragalactic Survey (JADES) Origins Fields, we report spectroscopic redshift measurements of \fsun{1,445} emission-line galaxies at $z=0-9$.
Within this sample, we identify two prominent galaxy protoclusters at $z = 3.47$ and 3.69, each anchored by massive dusty star-forming galaxies (DSFGs).
In the vicinity of these systems, we discover seven background galaxies at $z=3.6 - 6$ that simultaneously exhibit strong rest-frame optical emission lines (e.g., \oiii\ and \ha) and unusually reddened UV-to-optical continua.
We attribute this reddening to dust extinction arising from the circumgalactic medium (CGM) of the foreground DSFGs at projected separations of 7--30\,kpc.
We infer a high dust column density ($\gtrsim 10^{-1}$\,\msunsqpc), substantially exceeding those measured in low-redshift halos and those predicted by hydrodynamical simulations like IllustrisTNG and FIRE-2. 
The steep extinction curves, comparable to or steeper than that of the SMC, indicate a dominant population of small dust grains in the high-redshift CGM.
We conclude that DSFGs at this epoch host large reservoirs of dusty CGM enriched to solar metallicity.
These extended dust components are largely invisible to (sub-)millimeter interferometers such as ALMA because of their low surface brightness.
We discuss the physical processes in dust transport that might be key to reproducing our observations, including galaxy mergers, cool-phase gas outflows, dust shattering, sputtering and radiation pressure.
Finally, we caution that foreground CGM dust extinction may redden background galaxies at intermediate redshifts to mimic Lyman-break galaxies at $z\gtrsim10$.

\end{abstract}


\section{Introduction}
\label{sec:01_intro}

The baryon cycle is a fundamental driver of galaxy formation and evolution, regulating how gas is accreted, processed through star formation, and subsequently redistributed by feedback \citep[see reviews by][]{tumlinson17,peroux20,fco23}. Metals synthesized in stars and expelled into the circumgalactic medium (CGM) are key tracers of this cycle: they encode the integrated history of star formation, chemical enrichment, and outflow–inflow interactions. 

Among all metal constituents in the CGM, dust grains are especially informative. 
First, dust can be probed efficiently through its attenuation of background galaxies and quasars, providing a sensitive measure of the distribution of enriched material well beyond the stellar disk \citep[e.g.,][]{menard10, menard12, peek15, mccleary25, guha26}. 
Second, the grain size distribution and dust column density carry critical physical information about feedback-driven grain processing, including growth, shattering, sputtering, and other destruction mechanisms operating in the CGM environment \citep[e.g.,][]{hirashita21, hirashita24, otsuki24}. 
Together, dust and metals therefore offer a unique window into the exchange of material between galaxies and their surrounding halos, and into the physical mechanisms that regulate galaxy evolution.

Observational studies of dust and metal absorption in the CGM have historically been confined to low redshifts, largely because of the scarcity of sufficiently bright background sources and the challenges associated with obtaining high–signal-to-noise ultraviolet and optical spectra. 
Foreground CGM absorbers have therefore been traditionally probed using luminous background quasars, and large statistical samples have enabled constraints on dust reddening through stacking analyses \citep[e.g.,][]{menard10, mccleary25}.
Alternatively, using stacking analyses of millimeter data, \citet{meinke23} detected statistically significant dust emission extending to $\sim$Mpc scales around $z\sim1$ quiescent galaxies, providing new evidence for widespread dust in the CGM.

The advent of the James Webb Space Telescope \citep[JWST;][]{gardner23} has fundamentally transformed this landscape. Its infrared spectroscopic capabilities now permit deep photometric and spectroscopic observations of vast numbers of galaxies at high redshift, enabling systematic studies of CGM absorption well beyond the local Universe. 
In particular, NIRCam wide-field slitless spectroscopy (WFSS; \citealt{greene17, rieke23a_nrc}) provides unbiased spectroscopic sampling without pre-selection of targets and can capture spectra for closely separated galaxy pairs that would be difficult to observe with the NIRSpec micro-shutter assembly \citep[][]{jakobsen22, ferruit22}. 
This combination of depth, multiplexing, and spatial completeness offers an unprecedented opportunity to map CGM metals and dust across cosmic time.

In this work, we report the JWST discovery of unexpectedly large reservoirs of small dust grains in the CGM of massive star-forming galaxies at $z\sim 3.5$. 
This result emerges from the JWST Cycle 3 Guaranteed Time Observations (GTO) conducted as part of the JWST Advanced Deep Extragalactic Survey (JADES) Origins Fields (JOF; \citealt{eisenstein23_jades, eisenstein25_jof}), which provide deep NIRCam wide-field slitless spectroscopy over a $\sim$13-arcmin$^2$ region. 
From these data, we obtain spectroscopic redshifts for 1445 galaxies spanning a broad range of cosmic epochs, enabling an unprecedented census of line-emitting systems at high redshift.
This grism spectroscopic redshift catalog is publicly released through this work.
Within this sample, we identify a population of emission-line galaxies (ELGs) at $z = 3.6 - 6$ whose spectral energy distributions (SEDs) exhibit anomalously strong reddening. 
We demonstrate that this reddening is best explained by dust extinction arising from the halos of foreground massive star-forming galaxies at $z \sim 3.5$, with projected separations of 7--30 kpc. 

The structure of this paper is as follows. Section~\ref{sec:02_obs} describes the observations and data-processing procedures. 
Section~\ref{sec:03_res} presents the redshift measurements, sample definition, and SED modeling framework. 
Section~\ref{sec:04_disc} examines the physical interpretation of the inferred dust column densities and grain-size distributions, including comparisons with hydrodynamical simulations, ALMA observations and implications for the physics of dust transport from the ISM to the CGM. 
Section~\ref{sec:05_sum} summarizes our findings and outlines prospects for future work.
Throughout this paper we adopt a flat $\Lambda$CDM cosmology with $H_0 = 70$\,\si{km.s^{-1}.Mpc^{-1}} and $\Omega_M = 0.3$. 
All magnitudes are reported in the AB system \citep{oke83}.

\section{Observations and Data Processing}
\label{sec:02_obs}

\subsection{JWST and HST Imaging}
\label{ss:02a_img}

The JADES Origins Field (R.A.=53.06\arcdeg, Decl.=--27.87\arcdeg) is a $\sim9$-arcmin$^2$ deep field located in the southwest of the Great Observatories Origins Deep Survey South (GOODS-S) field \citep{giavalisco04}.
Ultra-deep JWST NIRCam imaging observations were obtained through JADES \citep{eisenstein23_jades} in Cycles 1--3, mostly as coordinated parallel observations of deep NIRSpec multi-object spectroscopic survey in the Hubble Ultra Deep Field \citep[HUDF;][]{beckwith06}.
We refer interested readers to \citet{eisenstein25_jof} for a detailed description of the observation descriptions and imaging data quality in the JOF.

In this work, we use the latest JADES DR5 data products, including imaging mosaics and photometric catalogs \citep[][]{johnson26, robertson26}.
These includes 16 JWST/NIRCam bands (F070W, F090W, F115W, F150W, F162M, F182M, F200W, F210M, F250M, F277W, F300M, F335M, F356W, F410M, F444W, F480M), five HST/ACS (F435W, F606W, F775W, F814W, F850LP) and four HST/WFC3-IR photometric bands (F105W, F125W, F140W, F160W) across 0.4--5.0\,\micron.
The typical $5\sigma$ point-source depth of NIRCam imaging data (measured with diameter $D = 0\farcs2$ aperture) reaches 30.6\,mag \citep{eisenstein25_jof}.
Archival HST data are taken from the Hubble Legacy Fields project (HLF v2.5; \citealt{whitaker19}) and matched to JADES astrometry.

\subsection{JWST/NIRCam Grism Spectroscopy}
\label{ss:02b_grism}

The JWST/NIRCam WFSS observations of JOF were obtained on UT October 15--20, 2024 through Cycle-3 GTO program 4540 (PI: Eisenstein).
Long-wavelength (LW) slitless spectroscopy with the column-direction grism (Grism C) were taken with three block filters, F322W2, F356W and F444W. 
Detailed descriptions of the observational design and implementation are presented by \citet{eisenstein25_jof}.

With tight 2$\times$2--pointing mosaics, the total survey area with complete spectral coverage across 2.4--5.0\,\micron, i.e., the complete NIRCam LW wavelength range, is 7.9\,arcmin$^2$.
The 4540 observations were designed to provide full spectral coverage at 3.0--5.0\,\micron\ for the ultra-deep JOF NIRCam imaging region (including the NIRCam module gaps) over 12.9\,arcmin$^2$.
The area with partial spectral coverage across the same wavelength range is up to 23.8\,arcmin$^2$.
The median $5 \sigma$ depth for unresolved emission lines reaches $0.9 \times 10^{-18}$\,\si{erg.s^{-1}.cm^{-2}} around 4.3\,\micron\ with the F444W filter at a median exposure time of 5.4\,hours.
The depths with the F322W2 and F356W filters alone, centered around 2.8 and 3.6\,\micron\ are $3.5\times10^{-18}$\,\si{erg.s^{-1}.cm^{-2}} and $1.0\times10^{-18}$\,\si{erg.s^{-1}.cm^{-2}}, respectively, and both have a median exposure time of 3.0\,hours. 
Note that both the F322W2 and F356W filters cover 3.1--4.0\,\micron, but we process the data separately because of the different flux calibration and spectral contamination in each band.

The NIRCam WFSS data were processed through the publicly available pipeline outlined by \citet{sun23}, with the most recent update on tracing function, wavelength calibration and modeled sky background by \citet{sunf25b}.
Customized steps beyond the standard JWST stage-1 reduction include flat-fielding, modeled sky background subtraction, 1/f noise subtraction, bad pixel removal, and astrometric correction.
2D and 1D spectral extraction was performed for all galaxies that are bright enough in the F277W, F356W and F444W bands (typically down to 29.6\,mag) that might yield $5\sigma$ emission-line detections with the grism data.
To produce data that are useful both for continuum science and emission-line identification, the spectra were extracted from the grism images both before and after continuum removal through iterative median filtering. 
1D spectra were extracted using both boxcar aperture (height\,=\,0\farcs31) and optimally  \citep{horne86} using the collapsed source profile.
We further performed automatic emission line identification on 1D spectra with S/N greater than 3 using various wavelength bin sizes (1--8\,nm; see \citealt{sunf24}).
This automatic emission line catalog is used later as priors for redshift identification (Section~\ref{ss:03a_z}).

\subsection{ALMA data}
\label{ss:02c_alma}

Through this work, we identified five dusty star-forming galaxies at $z\sim3.5$ that attenuate the background emission-line galaxies. 
To determine the centroids of their dust reservoirs in (sub-)millimeter continuum emission, we queried the ALMA data archive and found available data for three foreground galaxies. 
We therefore used the public archival ALMA continuum imaging data with lowest RMS noise available for each source, including Programs \#2015.1.00242.S (PI: Bauer; ALMA Band-7 at 870\,\micron) and \#2018.1.01079.S (PI: Franco; ALMA Band 4 at 2\,mm).
These archival data have been analyzed and reported in the literature including \citet{cowie18,zhoul20,mckay25}.

\section{Results}
\label{sec:03_res}

\subsection{Redshift Identification}
\label{ss:03a_z}

We measured the spectroscopic redshifts ($z_\mathrm{spec}$) of galaxies within the NIRCam WFSS footprint using the extracted spectra.
Because of the great depth and richness of medium bands, emission-line galaxies within the JOF feature highly accurate photometric redshifts ($z_\mathrm{phot}$) that have been validated by JADES NIRSpec surveys \citep{curtislake_dr4,deugenio25a_dr3,deugenio25d,scholtz25b_dr4}.
Specifically, the JADES ``Dark Horse'' pilot survey over the JOF suggests excellent \zph\ accuracy ($\langle (z_\mathrm{spec} - z_\mathrm{phot}) / (1 + z_\mathrm{spec}) \rangle = 0.011$) and very low outlier fraction ($f_\mathrm{outlier}=0.025$ above $5\sigma$; \citealt{deugenio25d}).
Therefore, to maximize the efficiency of redshift identification, we use the JADES \zph\ to compute the expected wavelength ranges of expected strong line detections (e.g., \oiii$\,\lambda\lambda$4960,5008, \ha, \siii\,$\lambda\lambda$9071,9533, \hei\,$\lambda$10833, \pab, \paa; cf.\ statistics of common emission lines in NIRCam WFSS data by \citealt{sunf25b}).
We then cross matched these with the automatic emission line catalog to tentatively assign redshift solutions. 
All redshift solutions are visually inspected in multiple iterations to correct for incorrect association because of spurious lines, contamination from other emitting galaxies on the spectral trace, and also the so-called ``tadpole'' ghost in NIRCam Grism C data that could mimic complex emission lines \citep{rigby23}.


In this work, we report the redshift identification of \fsun{1445} sources across $z = 0.1 - 8.7$ (median $z=3.47$).
The redshift catalog is further presented in Appendix~\ref{apd:01_z} and made publicly available through this work.
We caution that this redshift catalog is not complete, and thus careful statistics of the survey's completeness and galaxy luminosity function are subjects of future work.
Similar to \citet{sunf25b}, our redshift catalog consists of emission-line galaxies that (\romannumeral1) have at least one emission line detected at $\mathrm{S/N} \geq 4$, (\romannumeral2) have a quadratically combined S/N of all identified emission lines $\sqrt{\sum_i (\mathrm{S/N})_i^2}$ at $\geq 5$.
Note that the same spectral line observed with different filters is considered separately in the criteria above.
We measure the redshifts using the best-fit line centroids weighted by their line S/N at $\geq 4$.
Following the same criteria in \citet{linx25a} and \citet{sunf25b}, the confidence levels of \zsp\ are scored at 1--6 according to the number of line detections and $\delta z = z_\mathrm{spec} - z_\mathrm{phot}$, with \texttt{zconf=6} for the most confident redshifts determined by two line detections and $|\delta z| \leq 0.2$, and \texttt{zconf=1} for the least confident redshifts determined by single line and $|\delta z| \geq 1$.

\begin{figure*}[!t]
\centering
\includegraphics[width=\linewidth]{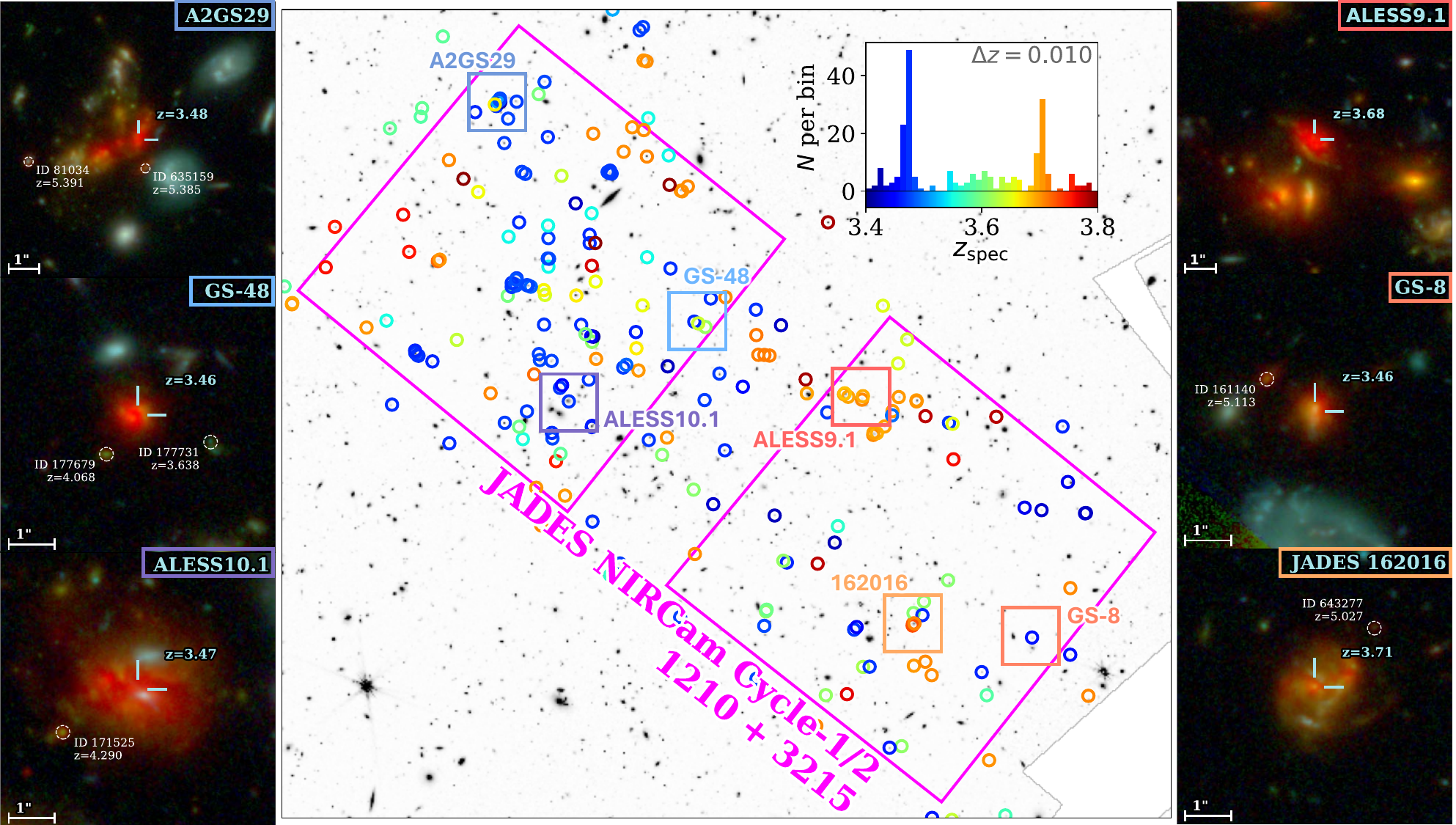}
\caption{Six dusty star forming galaxies spectroscopically confirmed within the $z=3.47$ and 3.70 protoclusters in the GOODS-S/JOF region.
The middle panel shows the on-sky distribution of galaxies with grism \zsp\ across 3.4--3.8 superimposed on NIRCam F444W image, with the inset panel displaying the redshift histogram (redshift color-coded).
The ultra-deep NIRCam imaging footprint in the JOF is outlined by the two 2\farcm2$\times$2\farcm2 magenta boxes.
The six image panels on the left and right show the NIRCam F444W-F200W-F115W RGB images of the selected DSFGs.
The ALMA centroids of the dust continuum emission are indicated by the skyblue ticks.
Seven reddened emission-line galaxies in the background of the five DSFGs (except for ALESS9.1) are highlighted by the dashed white circles with JADES NIRCam ID and \zsp\ indicated.
}
\label{fig:z3pc}
\end{figure*}

\begin{figure*}[!t]
\includegraphics[width=\linewidth]{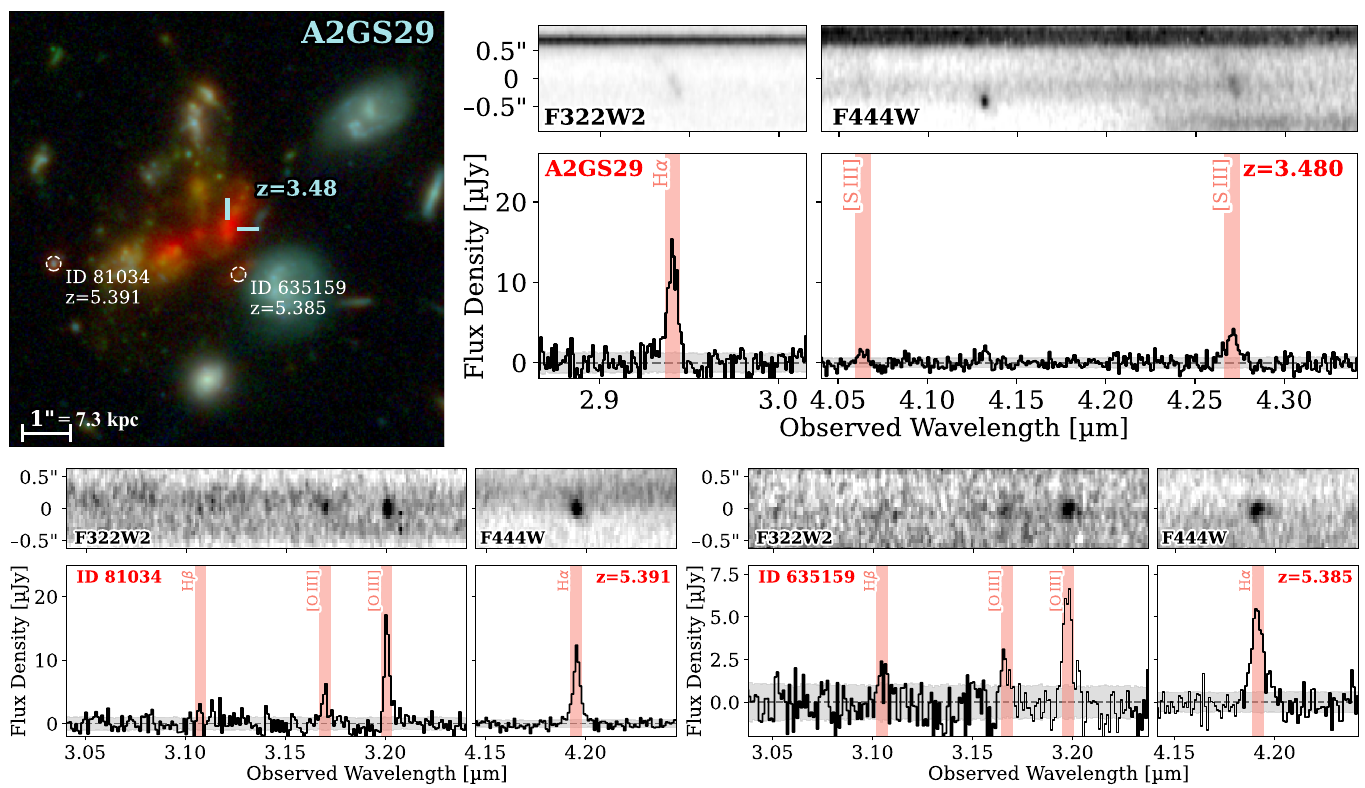}
\caption{JWST/NIRCam images and spectra of the massive DSFG A2GS29 and two reddened galaxies in the background. 
We show the F322W2 and F444W grism spectra in the wavelength ranges containing  \oiii, \ha\ and \siii\ with the emission lines highlighted.
The contaminated continuum is not subtracted in the 2D spectra, but is subtracted in the 1D spectra.
}
\label{fig:spec}
\end{figure*}

\subsection{Protoclusters at $z\sim3.5$ Anchored by DSFGs}
\label{ss:03b_pc}

Cycle-2 GO-3215 NIRCam imaging suggested the presence of an obvious overdensity of F300M-excess (relative to adjacent wide bands F277W and F356W) emission-line galaxies within the JOF \citep{eisenstein25_jof}.
Therefore, half of the originally planned LW grism observation time below 4\,\micron\ was split equally between F356W and F322W2 to cover the F300M bandpass (2.83--3.16\,\micron). 
This change proved successful: \fsun{226} \ha-emitting sources at $z_\mathrm{spec} = 3.4 - 3.8$ are identified over this wavelength range through our NIRCam WFSS survey.

Figure~\ref{fig:z3pc} shows the on-sky distribution and redshift histogram of these galaxies.
Two of the most prominent spikes are at $z=3.47$ and 3.70.
Each of the structures includes 86 and 53 sources, respectively, within a redshift offset $\Delta z = \pm 0.03$ and thus a line-of-sight velocity offset of $\Delta v \sim \pm 2000$\,\si{km.s^{-1}}. 
These two galaxy overdensities encompass $61\pm3\%$ of spectroscopically galaxies across $3.4<z<3.8$, and they have been reported by multiple works including \citet{frank16, guaita20, shah24}, and massive galaxies within the overdensities have been subject of follow-up studies with ALMA \citep[e.g.,][]{ginolfi17, zhoul20} and JWST/NIRSpec \citep[e.g.,][]{lamperti24, baker25, perna25a}.
Within our survey area ($5 \times 2$ comoving Mpc$^2$ at $z=3.5$), galaxies in the two overdensities outline filamentary structures with clear clumping around certain massive DSFGs with stellar mass $M_\mathrm{star} \sim10^{11}$\,\msun.
These DSFGs include (but not limit to) A2GS29 \citep[NIRCam WFSS $z=3.480$;][]{cowie18, franco18, zhoul20, gomez22}, ALESS10.1 \citep[NIRCam WFSS $z=3.472$, ALMA CO $z=3.467$;][]{hodge13,hodge25,zhoul20}, and ALESS9.1 (NIRCam WFSS $z=3.683$, ALMA CO $z=3.694$; \citealt{hodge13,hodge25,birkin21}; also known as ``Cosmic Rose''; \citealt{alberts24,zhuy25a_mb}).
These filamentary structures potentially extend towards the HUDF at $\sim10$ comoving Mpc away as suggested by MUSE surveys of \lya-emitting galaxies \citep{bacon21_od}.
Given the large number of confirmed member galaxies, extension of the filamentary structures and presence of vigorous starbursts (star-formation rate SFR\,$\gtrsim 10^3$\,\smpy), these two overdensities are comparable to protoclusters in simulations \citep[e.g.,][]{chiang17} at similar redshifts.

\begin{figure*}[!t]
\centering
\includegraphics[width=\linewidth]{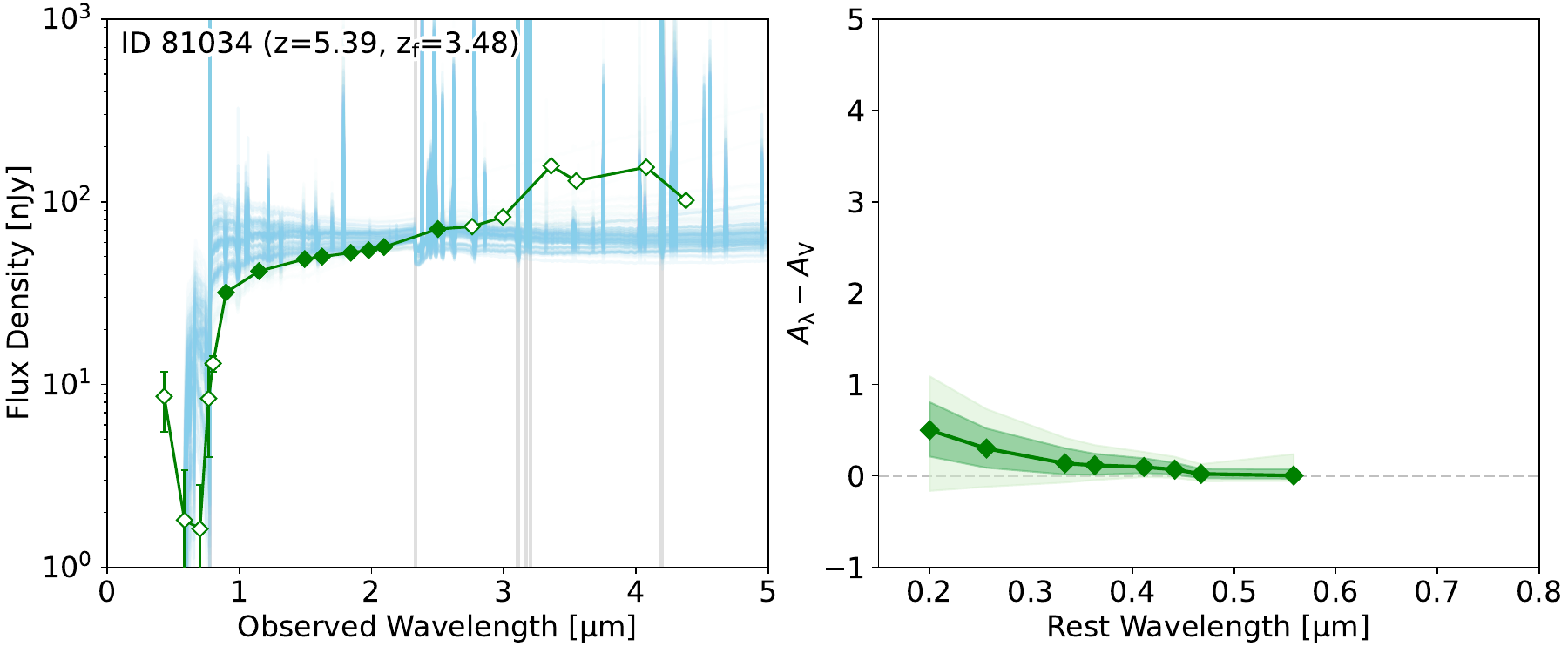}
\includegraphics[width=\linewidth]{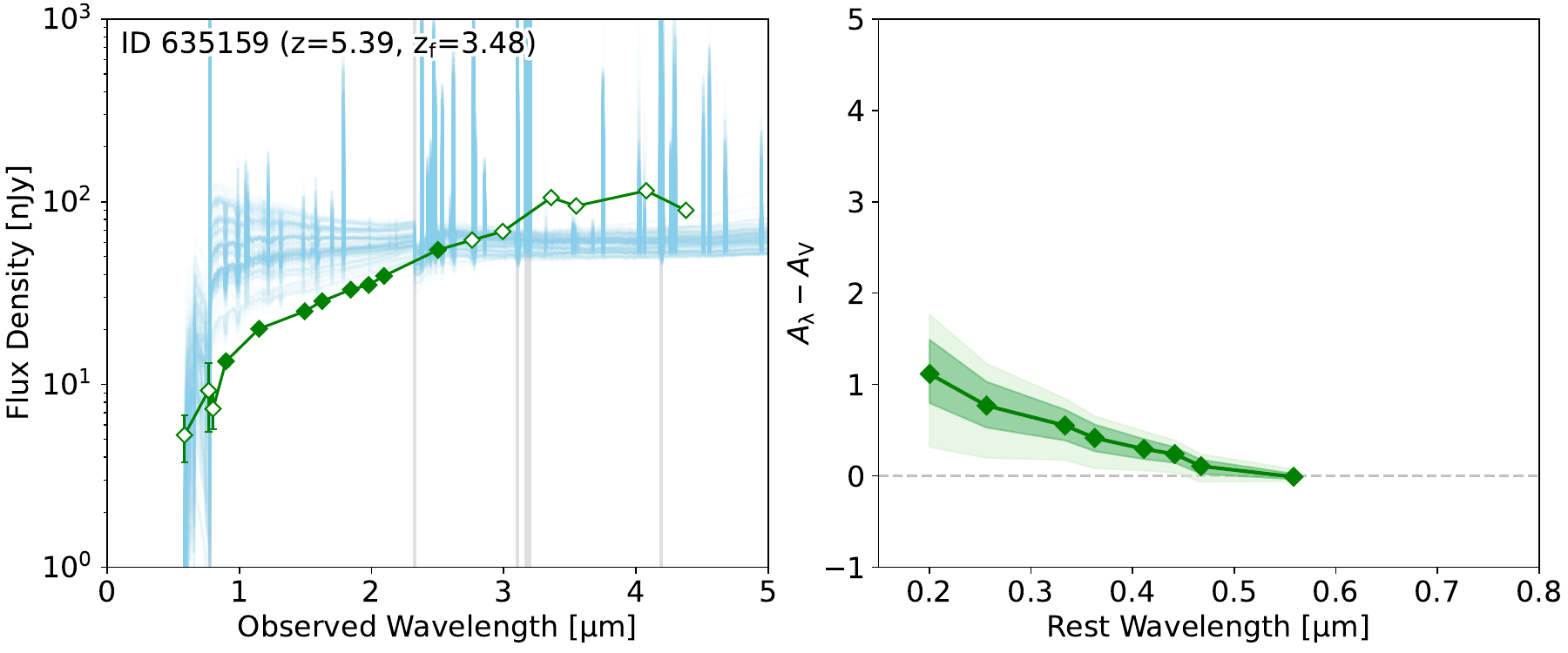}
\caption{\textbf{Left:} SEDs of two CGM-reddened galaxies behind A2GS29 (Figure~\ref{fig:spec}), shown as the green diamonds connected by green solid lines.
Open symbols denote the photometric bands affected by IGM attenuation or by the presence of strong lines (e.g., \hb, \oiii\ and \ha; wavelengths highlighted as vertical silver lines) within the bandwidths, and therefore are not used for extinction curve modeling.
The wavelength of the Balmer break is also indicated by a vertical silver line.
The SED templates  that fit other emission-line galaxies at similar redshifts and brightness but without foreground CGM extinction are shown as the skyblue lines.
These templates are projected to the same redshifts of the CGM-reddened galaxies and normalized by the flux density at 5\,\micron.
\textbf{Right:} inferred dust extinction curve ($A_\lambda - A_V$; see Section~\ref{ss:03d_ext}) from the foreground CGM, shown as the filled green diamonds connected by the solid lines.
$2\sigma$ and $1\sigma$ confidence intervals are shown as the light and dark green-shaded regions, respectively.
}
\label{fig:sed_a2gs29}
\end{figure*}

\subsection{Reddened Emission-Line Galaxies}
\label{ss:03c_red}

In the vicinity of the aforementioned DSFGs (Figure~\ref{fig:z3pc}), we identify a population of galaxies in the background ($z_\mathrm{spec} = 3.6 - 6$) with strong rest-frame optical emission lines (e.g., \oiii\ and \ha) and reddened UV--optical continuum emission.
Figure~\ref{fig:spec} shows two examples of such systems in the background of A2GS29.
The redshifts of the foreground DSFGs and background emission-line galaxies are all confirmed at \texttt{zconf=6} through the detection of multiple lines including \hb, \oiii\,$\lambda\lambda$4960,5008, \ha\ and \siii\,$\lambda\lambda$9071,9533.
Notably, these two sources in the background are members a known galaxy protocluster at $z=5.39$ in the GOODS-S field \citep{helton24a}.

The spectral energy distributions (SEDs) of the two background sources are shown in Figure~\ref{fig:sed_a2gs29} (left panels) as the green diamonds and solid lines.
For comparison, the SEDs of emission-line galaxies within our spectroscopic sample at similar redshifts ($3.5 < z_\mathrm{spec} < 6.5$) and brightness ($26.0 < \mathrm{F444W\,[mag]} < 29.0$) are shown, normalized to the same redshifts (blue circles connected with lines).
It is obvious that the two background sources (JADES NIRCam ID 81034 and 635159) appear redder in the overall rest-frame UV--optical SEDs than most of sources used for comparison.

\begin{figure*}[!t]
\centering
\includegraphics[width=\linewidth]{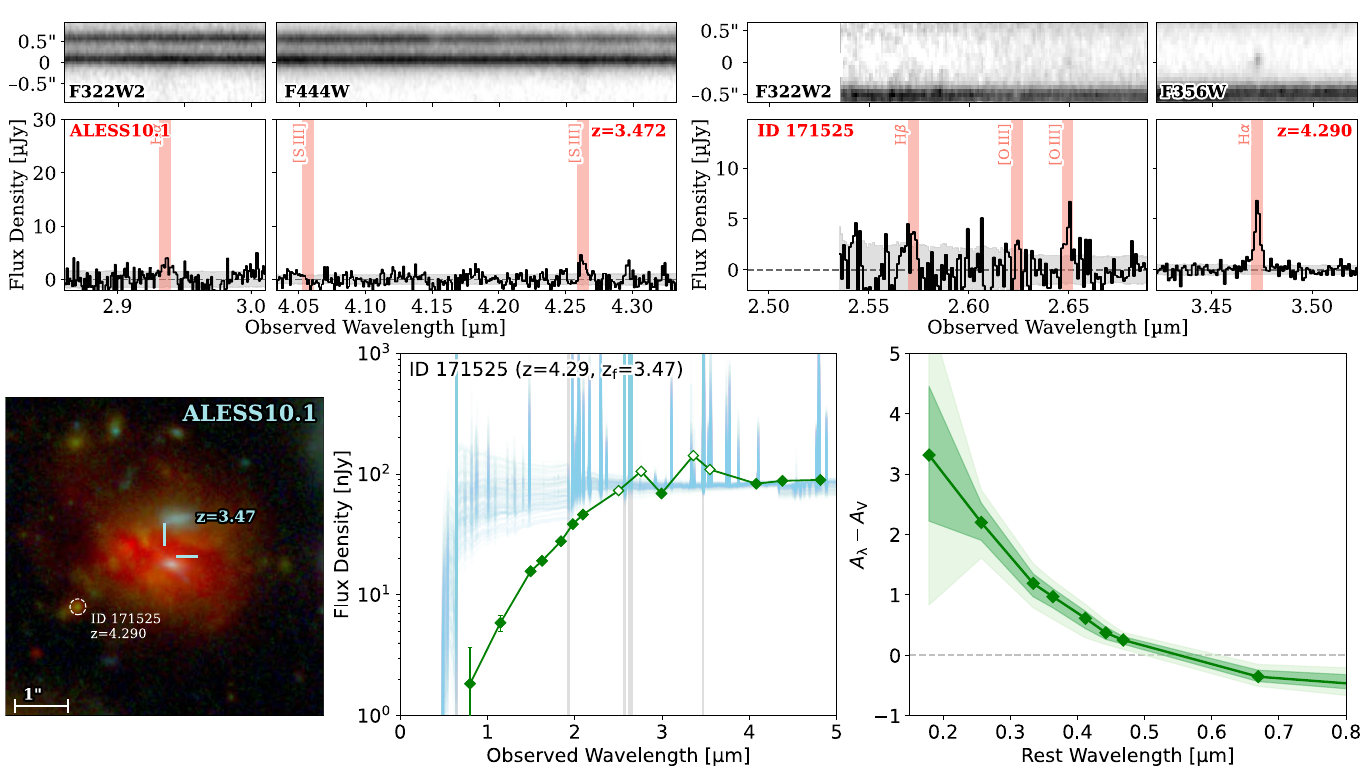}
\caption{Heavy dust extinction seen in JADES ID 171525 from the CGM of foreground DSFG ALESS 10.1. 
Top: NIRCam grism 2D and 1D spectra of foreground ALESS10.1 and background ID 171525 (similar to that of Figure~\ref{fig:spec}).
Bottom left: NIRCam image of the system (same as that in Figure~\ref{fig:z3pc}). 
Bottom middle/right: SED and inferred extinction curve seen along the line-of-sight of 171525 (similar to that of Figure~\ref{fig:sed_a2gs29}). 
}
\label{fig:aless10_1}
\end{figure*}

A similar effect is observed for other systems of background galaxies with foreground DSFGs. 
The strongest effect is seen for ALESS10.1 and the background galaxy ID 171525 ($z=4.290$), highlighted in Figure~\ref{fig:aless10_1}.
With strong \ha\ and \oiii\,$\lambda$5008 equivalent widths respectively of 367$\pm$20\,\AA\ and 339$\pm$71\,\AA\ that are typical of star-forming galaxies at this epoch \citep[e.g.,][]{faisst16,sunf25b}, ID 171525 exhibits a very red UV--optical SED that cannot be explained by the dust attenuation within its own interstellar medium (ISM) or CGM.
Qualitatively, such heavy dust extinction is frequently seen in very massive galaxies (stellar mass $M_\mathrm{star} \gtrsim 10^{10}$\,\msun) with high dust mass ($M_\mathrm{dust} \gtrsim 10^{8}$\,\msun) and high dust surface density ($\Sigma_\mathrm{dust} \gtrsim 10$\,\msunsqpc), but atypical of low-mass ($M_\mathrm{star} \lesssim 10^{9}$\,\msun), young galaxies with strong emission lines as we observed.
The \ha-to-\hb\ flux ratios of these background galaxies are also consistent with the expected ratio for Case-B recombination without dust extinction (2.86; \citealt{osterbrock06}) despite a relatively large \hb\ flux uncertainty.
We also check the aperture used for photometry (diameter $D=0\farcs3$) and conclude that there is no significant contamination from either the foreground DSFG or other sources.
Therefore, a natural explanation for the reddened background SED is the additional dust extinction from the CGM of foreground DSFGs.


\subsection{Dust Extinction Modeling}
\label{ss:03d_ext}

We quantify the strength of foreground dust extinction through the modeling of the observed SEDs of background galaxies.
We caution that the background galaxies are both attenuated by the dust in their birth clouds, ambient ISM and potentially CGM at their redshifts (i.e., in situ), and also extincted by a slab of dust in the CGM of foreground DSFGs at lower redshifts.
The dust attenuation and extinction at the two redshifts are expected to be strongly degenerate. 
To bypass the degeneracy in fitting two dust components simultaneously, we do not conduct direct physical SED modeling of the CGM-reddened galaxies.
Instead, we assume that the CGM-reddened background galaxies have intrinsic SEDs similar to some of the field galaxies without foreground extinction. 
We therefore (\romannumeral1) construct intrinsic SED templates for the background galaxies using other emission-line galaxies in our NIRCam WFSS sample; (\romannumeral2) fit the required dust extinction curve to match the SED templates with the observed SEDs of CGM-reddened galaxies.

We obtain physical SED modeling of 429 galaxies at $3.5 < z_\mathrm{spec} < 6.5$ selected in Section~\ref{ss:03a_z} using the \textsc{cigale} \citep[][]{noll09, cigale} software.
To fit the 25-band photometry with HST/ACS, WFC3-IR and JWST/NIRCam at 0.4--5.0\,\micron, we assume a commonly adopted delay-$\tau$ star-formation history ($\mathrm{SFR}[t] \propto t \exp[-t / \tau]$) with optional late-time starburst, the \citet{bc03} stellar population synthesis models, and allow a metallicity range of $0.2 Z_\mathrm{\odot} \leq Z  \leq Z_\mathrm{\odot}$ and ionization parameter $-4.0 \leq \log U_\mathrm{ion} < -1.0$.
We adopt a modified \citealt{calzetti00} attenuation law, allowing the variation of the power-law slope by $\pm 0.3$, nebular gas emission line color excess $E(B - V)_g \leq 0.6$ and a reduction factor to be applied on $E(B - V)_g$ to derive the stellar continuum $E(B - V)_s$ as $0.4 - 0.7$, following the observations of dusty star-forming galaxies at similar redshifts \citep{sunf25a,tsujita25}.
The best-fit SED models of all galaxies are collected and adopted as intrinsic templates for the SED modeling of the CGM-reddened galaxy.

\begin{figure*}[!t]
\centering
\includegraphics[width=0.49\linewidth]{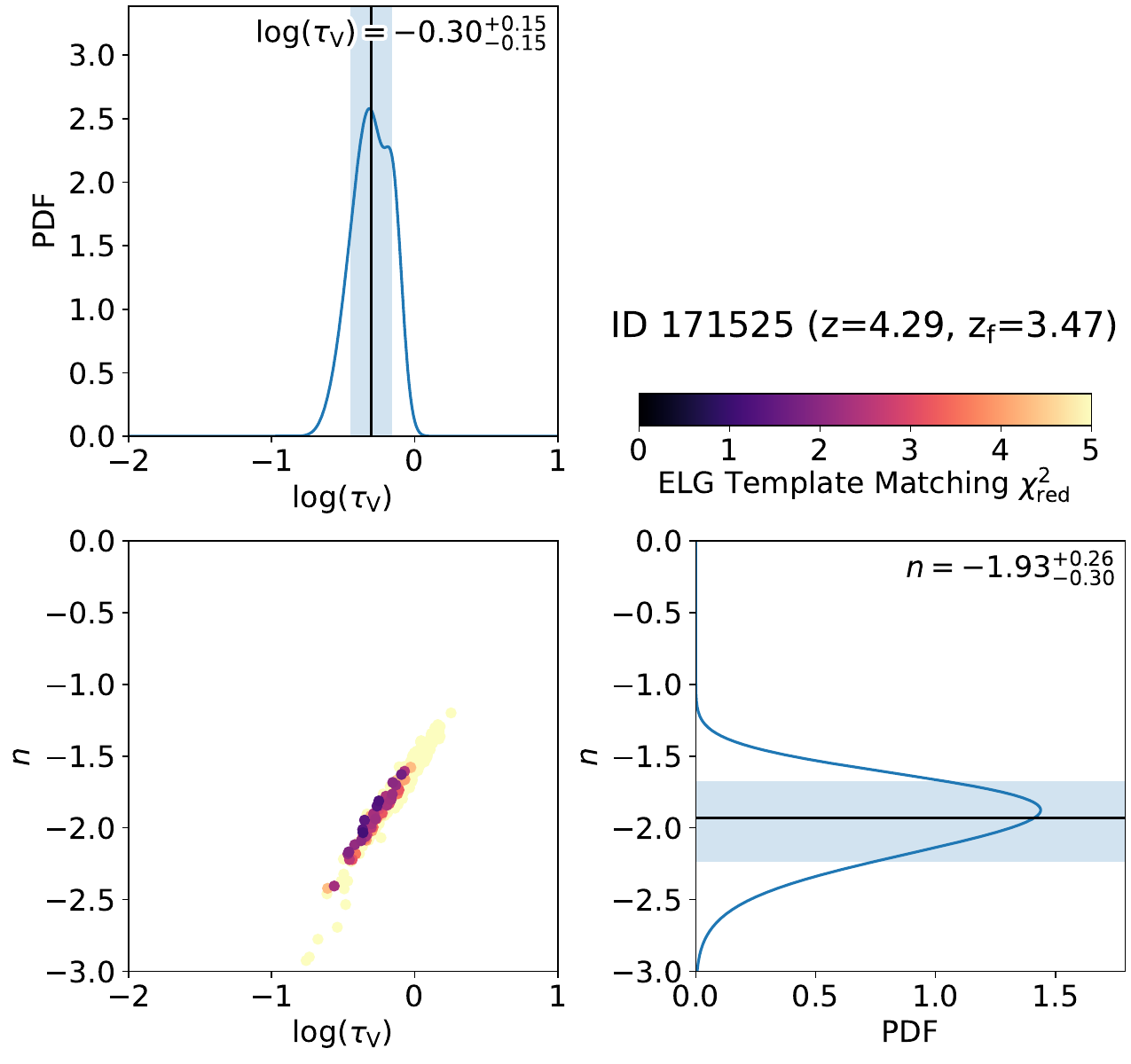}
\includegraphics[width=0.49\linewidth]{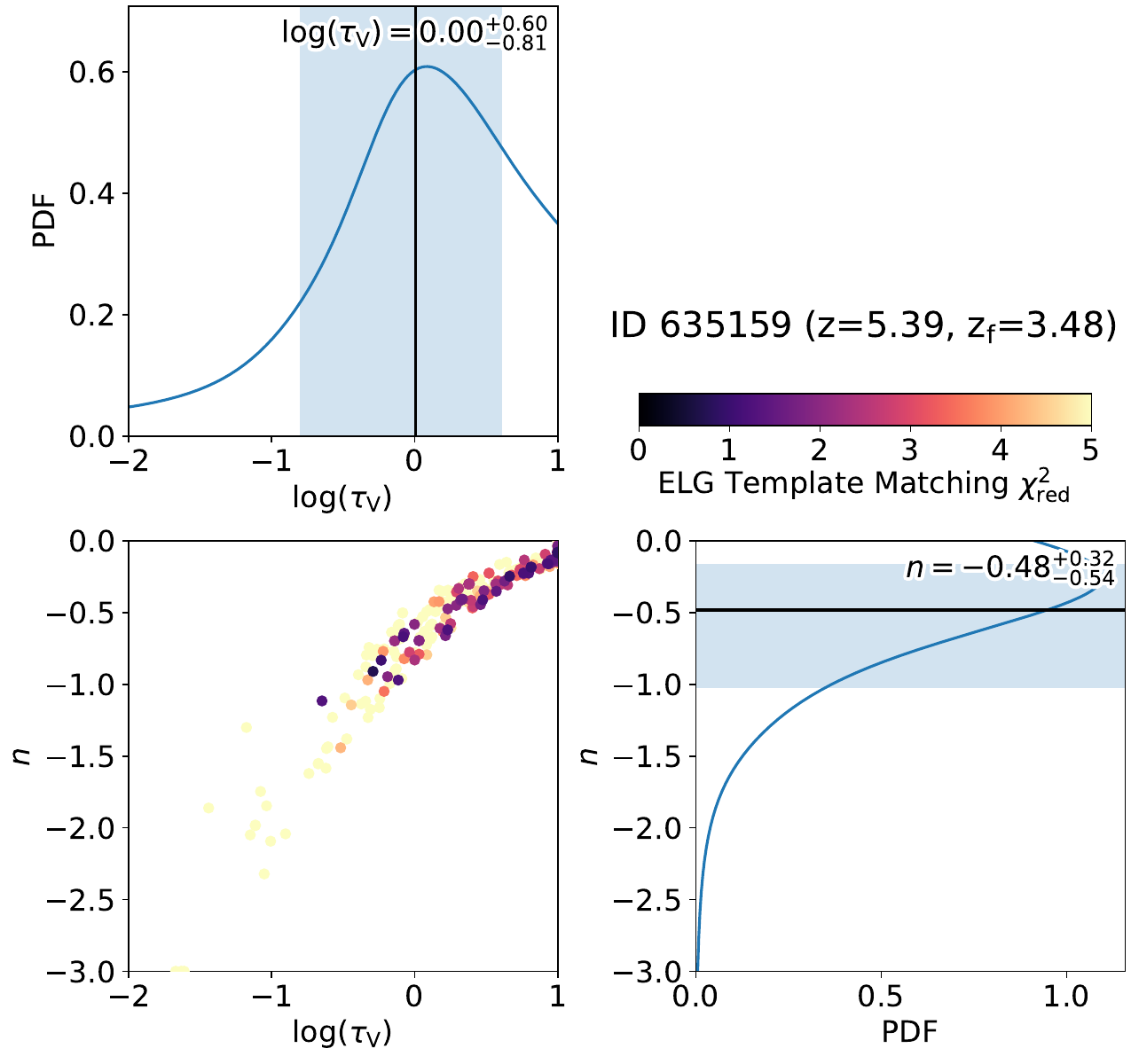}
\caption{Probability distribution of V-band optical depth of dust extinction $\log(\tau_V)$ and power-law stope $n$ of the extinction curve along the sightlines of two background ELGs (left: ID 171525 in the background of ALESS10.1 shown in Figure~\ref{fig:sed_a2gs29}; right: ID 635159 in the background of A2GS29, shown in Figure~\ref{fig:aless10_1}).
Each circle denotes the best-fit $\log(\tau_V)$ and $n$ using each spectral templates, color-coded by the reduced $\chi^2$ of template matching. 
Posterior distributions of $\log(\tau_V)$ and $n$, including the 16--50--84th percentiles of the credible intervals are shown as the skyblue bands and the black solid lines.
}
\label{fig:tau_vs_n}
\end{figure*}

\begin{deluxetable*}{@{\extracolsep{0pt}}ccccccccccccc}[!th]
\tablecaption{Inferred properties of the CGM dust extinction from background emission-line galaxies.}
\label{tab:01_ext}
\tablewidth{0pt}
\tabletypesize{\footnotesize}
\tablehead{
\colhead{ID} & \colhead{R.A.} & 
\colhead{Decl.} & \colhead{$z_\mathrm{spec}$} & \colhead{\texttt{zconf}} & \colhead{F444W} & \colhead{Foreground} & \colhead{R.A.$_f$} & 
\colhead{Decl.$_f$}  & \colhead{$z_f$} & \colhead{$R$} & \colhead{$\log(\tau_V)$} &  \colhead{$n$} \\
\colhead{} & \colhead{[deg]} & \colhead{[deg]} & \colhead{} & \colhead{} & \colhead{[mag]} & \colhead{} & \colhead{[deg]} & \colhead{[deg]} & \colhead{} & \colhead{[\arcsec]} & \colhead{} & \colhead{} 
}
\startdata
 81034  & 53.08832 & $-27.84042$ & 5.391 & 5 & 26.4  & A2GS29     & 53.08719 & $-27.84023$ & 3.48 & 3.66 & $-0.47_{-0.62}^{+0.65}$ & $-1.07_{-0.73}^{+0.65}$ \\
161140  & 53.02564 & $-27.89496$ & 5.113 & 4 & 26.6  & GS-8       & 53.02531 & $-27.89515$ & 3.46 & 1.26 & $-0.04_{-0.22}^{+0.21}$ & $-1.19_{-0.36}^{+0.31}$ \\
171525  & 53.07992 & $-27.87106$ & 4.290 & 4 & 26.5  & ALESS10.1  & 53.07941 & $-27.87080$ & 3.47 & 1.89 & $-0.30_{-0.15}^{+0.15}$ & $-1.93_{-0.30}^{+0.26}$ \\
177679  & 53.06499 & $-27.86287$ & 4.068 & 1 & 27.3  & GS-48      & 53.06478 & $-27.86263$ & 3.46 & 1.09 & $-0.30_{-0.26}^{+0.21}$ & $-1.65_{-0.38}^{+0.29}$ \\
177731  & 53.06428 & $-27.86279$ & 3.638 & 6 & 28.1  & GS-48      & 53.06478 & $-27.86263$ & 3.46 & 1.70 & $-0.96_{-0.41}^{+0.38}$ & $-1.79_{-0.58}^{+0.42}$ \\
635159  & 53.08711 & $-27.84049$ & 5.385 & 5 & 26.5  & A2GS29     & 53.08719 & $-27.84023$ & 3.48 & 0.97 & $+0.00_{-0.81}^{+0.60}$ & $-0.48_{-0.54}^{+0.32}$ \\
643277  & 53.03879 & $-27.89336$ & 5.027 & 6 & 28.5  & 162016     & 53.03920 & $-27.89372$ & 3.71 & 1.83 & $-0.32_{-0.34}^{+0.32}$ & $-1.58_{-0.67}^{+0.65}$ \\
\enddata
\tablecomments{Subscript $_f$ indicates the information for the foreground DSFG, including coordinates and redshifts. $R$ is the projected separation between the centroids of foreground DSFG (measured using ALMA dust continuum if available) and background galaxies, i.e., impact parameter, in units of arcsec. 1\arcsec\ corresponds to 7.3\,kpc at $z=3.5$.
$n$ is the power-law index of the extinction curve.
Redshifts are from NIRCam WFSS observations with typical uncertainty of $\Delta z = 0.001$.
The redshift uncertainty could be larger for the foreground DSFGs (up to  $\Delta z \sim 0.01$) because of the complex geometry between the \ha-emitting regions and continuum (or mass) centroids (see Section~\ref{ss:03b_pc}).
}
\end{deluxetable*}

We assume a simple power-law extinction curve with optical depth $\tau (\lambda) = \tau_V (\lambda / 0.55) ^ n$, where $\lambda$ is the rest-frame wavelength in units of micron, $\tau_V$ is the V-band optical depth, and $n$ is the power-law index.
Through our fitting, we first project the SED templates to the redshifts of CGM-reddened galaxies and compute the flux density ratios between the observed SEDs ($F_\mathrm{obs}$) and templates ($F_\mathrm{tmpl}$) in all bands.
The flux density ratios are therefore modeled as:
\begin{equation}
\ln[\frac{F_\mathrm{obs}}{F_\mathrm{tmpl}}
(\lambda_\mathrm{rest})] = C + \tau_V [1 - (\lambda_\mathrm{rest} / 0.55)^n]
\label{eq:fit_ext}
\end{equation}
where $C$ is the normalization in rest-frame $V$ band. 
Note that wavelength $\lambda_\mathrm{rest}$ refers to the rest frame of the foreground CGM, which is at redshift $z_\mathrm{f}$.
In practice, we fit $\log(\tau_V)$ to enforce $\tau_V > 0$, and we allow $-2 \leq \log(\tau_V) \leq 1$ and $-3 \leq n \leq 0$ that are reasonable from the data.
We do not use the photometric bands that are affected by the presence of strong \hb, \oiii, \ha\ lines, and intergalactic medium (IGM) attenuation. 
The least-squares fitting procedures and examples of results are shown in Figure~\ref{fig:sed_a2gs29} and Figure~\ref{fig:aless10_1}. 
All SEDs are projected to the observed frame for the background galaxies, and for a fair comparison, the spectral templates are normalized by Equation~\ref{eq:fit_ext} at 5.0\,\micron\ in the observed frame.
The logarithmic flux density ratios $F_\mathrm{obs} / F_\mathrm{tmpl}$ subtracted by $C$ can be directly translated to $A_\lambda - A_V$. 
We observe a clear trend of increasing dust extinction at bluer wavelengths in the foreground CGM.

Our fitting method cannot fully resolve the degeneracy between foreground CGM dust extinction and in-situ dust attenuation.
As shown in Figure~\ref{fig:tau_vs_n}, the best-fit $\log(\tau_V)$ and $n$ from each template matching show clear degeneracy, driven by the intrinsic color of the spectral template.
We caution that the majority of spectral templates may not match perfectly to the intrinsic continuum of the CGM-reddened galaxies.
Therefore, we only adopt the best 5\% of the templates (resultant $N \sim 20$) with smallest reduced $\chi^2$ to match the observed CGM-reddened SEDs.
We then derive the posterior distributions of $\log(\tau_V)$ and $n$ by combining the posterior of each best-fit result with equal weight.
Including more templates (e.g., the best 10\%) does not change the derived posteriors significantly.
Table~\ref{tab:01_ext} summarizes the 16-50-84th percentiles of $\log(\tau_V)$ and $n$ for each background source.

\section{Discussion}
\label{sec:04_disc}

\begin{figure*}[!t]
\centering
\includegraphics[width=\linewidth]{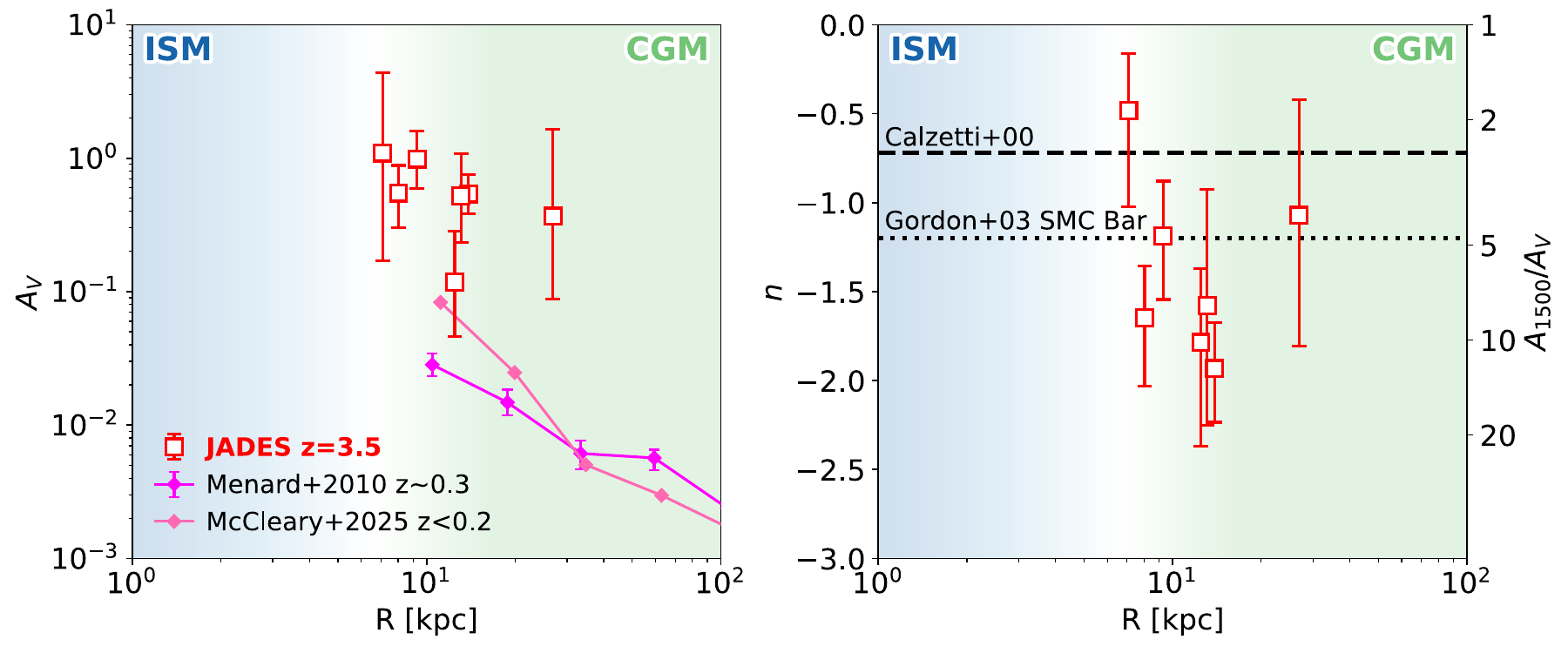}
\caption{\textbf{Left}: CGM dust extinction in the rest-frame $V$ band versus impact parameter. 
Our measurements at $z\sim3.5$ are highlighted as open red squares, and for comparison we show the CGM dust extinction measured by \citet{menard10} at $z\sim0.3$ and \citet{mccleary25} at $z<0.2$.
\textbf{Right}: Slope of extinction curve versus impact parameter. 
For comparison we draw the characteristic slopes of the \citet{calzetti00} attenuation curve ($n \sim - 0.7$; dashed gray line) and the \citet{gordon03} SMC bar extinction curve ($n \sim - 1.2$; dotted gray line) over $0.2 \sim 0.5$\,\micron.
The radial transition from ISM to CGM at $R\sim10$\,kpc is color-coded in the background of both panels.
}
\label{fig:av_n_r}
\end{figure*}

\subsection{Large Reservoirs of Dust in CGM}
\label{ss:04a_colden}

The left panel of Figure~\ref{fig:av_n_r} presents the rest-frame V-band extinction ($A_V$) versus the projected separation between the centroids of foreground DSFGs and background galaxies (i.e., impact parameter) measured with JADES.
Our observations suggest that the $z\sim3.5$ DSFGs in our sample produce significant extinction (up to $A_V \sim 1$) and reddening to the background galaxies through dusty CGM.
Such an extinction is observed even out to an impact parameter of 27 proper kpc as observed for ID 81034 in the background of A2GS29 (Figure~\ref{fig:sed_a2gs29}), although we caution that the uncertainty with this measurement is relatively large ($\Delta\log(\tau_V) \sim 0.6$).

We do not observe any obvious dependence of $A_V$ on impact parameter, although this is likely because of a small sample size.
We do not find any robust CGM-reddened galaxies at larger impact parameter ($R > 30$\,kpc), indicating that high-redshift CGM dust extinction is negligible at such a distance.
However, within $R < 30$\,kpc from the five foreground DSFGs, our sample is complete for the background emission-line galaxies above the WFSS detection limit, and all background sources are found to be reddened by foreground CGM dust. 
Detecting extinction along seven out of seven sightlines implies a high covering fraction ($f_c > 0.65$, $2\sigma$ confidence level assuming a binomial distribution) of CGM dust within $\sim 30$\,kpc of these foreground DSFGs.
Such a high covering factor is reminiscent of the near-unity covering factor of \mgii\ absorbers observed for $z\lesssim1$ galaxies once intervening background quasar sightlines at similar impact parameter \citep[$R \lesssim 20$\,kpc; e.g.,][]{nielsen13, farina14, guha26}. 

We also compare the observed $A_V$ with measurements in the literature at low redshifts.
At $z \lesssim 0.3$, the dust extinction from the CGM of foreground galaxies has been measured using background quasars \citep{menard10} and luminous red galaxies (LRGs; \citealt{peek15, mccleary25}) at impact parameters of 10\,kpc to 10\,Mpc.
As compared in the left panel of Figure~\ref{fig:av_n_r}, the CGM dust extinction of JADES DSFGs at $z\sim3.5$ is stronger than those of low-redshift galaxies selected with ground-based all-sky surveys by a factor of $\sim10$.

\subsection{Dust Grain Size and Composition}
\label{ss:04b_grain}

\begin{figure*}[!t]
\includegraphics[width=\linewidth]{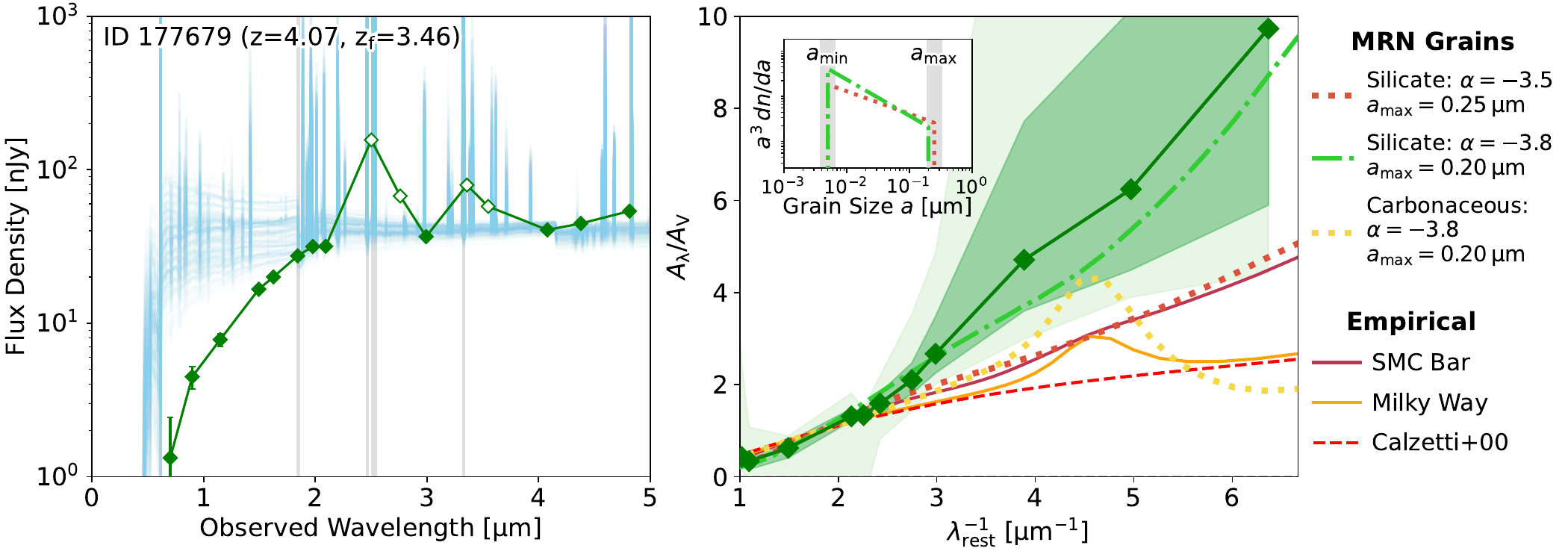}
\caption{Observed SED of ID 177679 behind GS-48 (left panel; similar to Figure~\ref{fig:sed_a2gs29}) and the inferred CGM dust extinction curve (right panel; filled green diamonds connected with solid lines).
$2\sigma$ and $1\sigma$ credible intervals of the extinction curve, in terms of $A_\lambda / A_V$, are drawn as the light and dark green-shaded regions, respectively.
For comparison we draw the extinction curve of the SMC bar \citep{gordon03}, the Milky Way average \citep{gordon09} and \citet{calzetti00} attenuation curves.
Extinction curves of silicate and carbonaceous dust grains following power-law \citep[][MRN]{mrn77} grain size distributions are also shown for comparison.
The inset panel presents two examples and parameterizations of grain size distributions that are drawn in the right panel for comparison.
}
\label{fig:ext_curve}
\end{figure*}

The shape of the dust extinction curve carries key information on the CGM dust grain size and composition.
The right panel of Figure~\ref{fig:av_n_r} shows the best-fit power-law slope $n$ of the extinction curve versus impact parameter.
Although we do not observe any significant dependence of $n$ on impact parameter, we find that most sightlines suggest a very steep slope for the extinction curve (median $n = -1.5 \pm 0.2$).
For comparison, the slope of the SMC bar extinction curve \citep{gordon03} modeled through a simple power law across 0.15--0.55\,\micron\ is $n \sim -1.2$, and the slope of the \citet{calzetti00} starburst galaxy attenuation curve is $n \sim -0.7$.
Five out of seven sightlines deviate from the \citet{calzetti00} slope at $>1\sigma$ significance, and the extinction curve sightlines along 171525 (Figure~\ref{fig:aless10_1}) and 177679 (Figure~\ref{fig:ext_curve}) are also steeper than that of SMC bar at $>2\sigma$ significance.

Such a steep extinction curve suggests very small dust grains in the CGM, i.e., on average similar to or even smaller than those in the SMC bar \citep[e.g.,][]{pei92, wd01, gordon03}.
Similar effects have been observed through the reddening curves of background quasars by foreground \mgii\ absorbers at $z\simeq1-2$ \citep{menard12}.
In addition to this, our extinction curve measurements cover the 2175-\AA\ bump for three of the sources at $z<4.3$. 
As highlighted in Figure~\ref{fig:ext_curve} through comparisons with the empirical extinction curves of the SMC bar \citep{gordon03}, Milky Way average \citep{gordon09}, and \citet{calzetti00} attenuation curves, we do not detect any conspicuous 2175-\AA\ bump in the CGM extinction curve, but instead a continuously rising $A_\lambda$ curve from 1 to 0.15\,\micron.
This is an interesting contrast to the prominent 2175-\AA\ bump seen in the ISM attenuation curve of certain high-redshift galaxies observed through JWST/NIRSpec \citep[e.g.,][]{witstok23}.

We interpret the observed extinction curve as being caused by a dominant population of small silicate dust grains.
Figure~\ref{fig:ext_curve} also shows the extinction curves computed using the \textsc{DGFit} package\footnote{\url{https://github.com/karllark/DGFit}} (K.\ Gordon \& K.\ Misselt, in prep.) assuming the silicate and carbonaceous extinction cross sections from \citet{laor93} and \citet{li01}.
We assume the MRN \citep{mrn77} power-law dust grain size distribution $dn/da \propto a^\alpha$ with cutoffs at $a_\mathrm{min}$ and $a_\mathrm{max}$ (Figure~\ref{fig:ext_curve} inset panel).
Here, $dn$ is the number density of dust grains over a small grain size bin $(a, a + da)$.
We adopt $a_\mathrm{min}$ at 0.005\,\micron\ \citep{mrn77, draine84}.
The commonly adopted parameters $\alpha = -3.5$ and $a_\mathrm{max} = 0.25$\,\micron\ with silicate grains can reproduce the steep extinction curve as observed in the SMC bar, while carbonaceous grains feature a shallower underlying slope over the UV wavelength range and a prominent 2175-\AA\ bump.
To reproduce the ultra-steep extinction curve seen with the 177679 sight line, it is necessary to increase the fraction of small dust grains at $a \lesssim 0.05$\,\micron. 
This can be achieved by further reducing the maximum size cutoff (e.g., $a_\mathrm{max} \sim 0.2$\,\micron) and increasing the power-law slope of the MRN grain size distribution (e.g., $\alpha = -3.8$) shown as the dash-dotted lime green curve in Figure~\ref{eq:fit_ext}.
For comparison, the extinction curve from carbonaceous dust grains with the same grain size distribution (dotted yellow line) clearly deviates from our observations, suggesting a minor contribution to the CGM dust extinction.

The discovery of a CGM dust composition dominated by silicate rather than carbonaceous grains is intriguing. 
At metallicities typical of DSFGs ($\sim Z_\odot$), low-mass asymptotic giant branch (AGB) stars ($M \lesssim 3.5$\,\msun) produce predominantly carbonaceous dust, whereas silicate dust production is expected to dominate at higher AGB masses ($M \simeq 3.5 - 8$\,\msun; e.g., \citealt{Zhukovska08, Ventura12b, Nanni14, sm24}). 
In contrast, core-collapse supernovae (CCSNe), such as SN 1987A, produce dust spanning a wide range of compositions. 
At small grain sizes ($a \sim 10^{-2}$\micron), the dust population is predicted to be dominated by silicates \citep{sarangi15, sluder18, sm24}. 
Taken together, the predominance of small silicate grains in the CGM suggests an origin linked to massive stars--through high-mass AGB evolution and/or CCSNe--associated with intense star formation during the early evolutionary phases of DSFGs.
We further discuss the physical implication for dust grain transport in Section~\ref{ss:04e_phys}.

\begin{figure*}[!t]
\centering
\includegraphics[width=\linewidth]{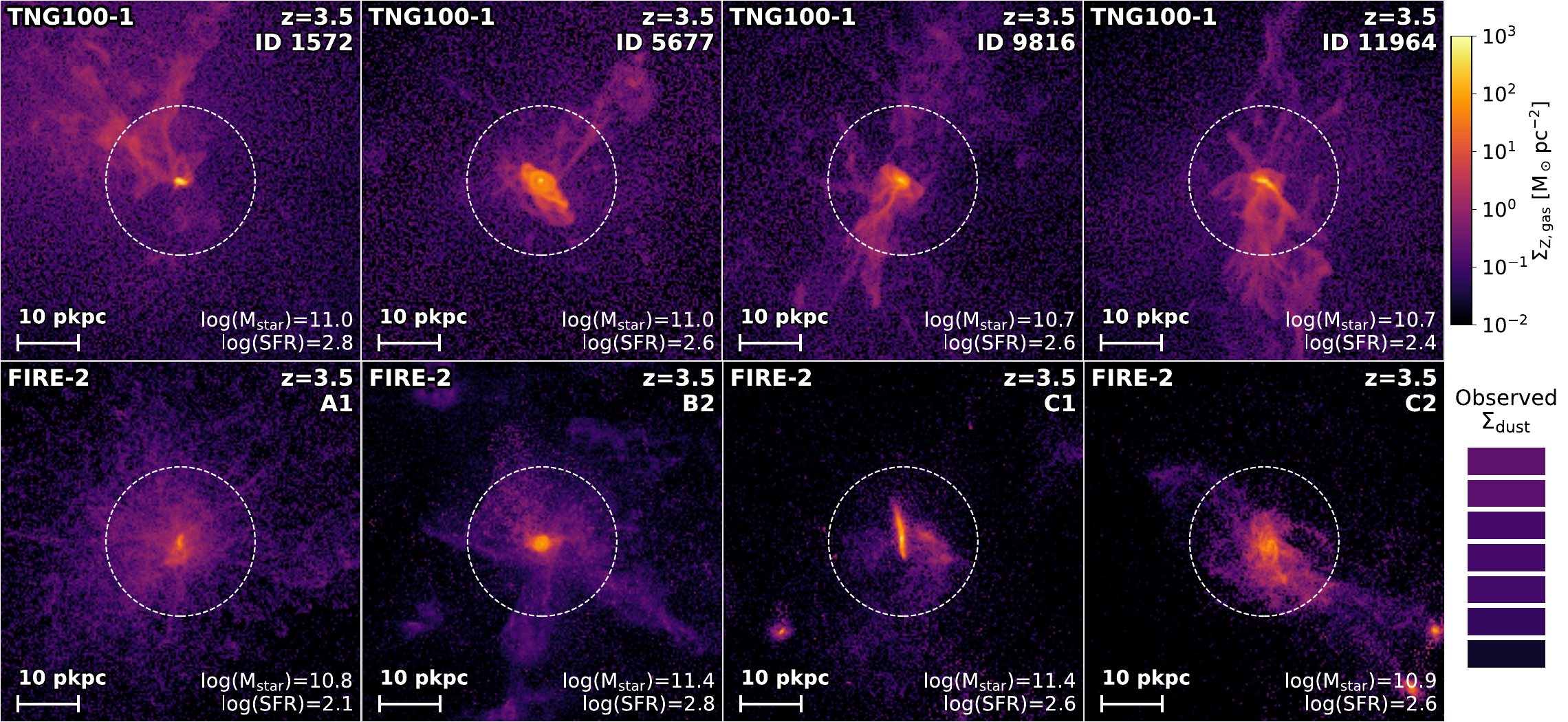}
\caption{Surface density maps of gas-phase metals ($\Sigma_\mathrm{Z, gas}$) in the halos of massive star-forming galaxies at $z = 3.5$ from TNG100 (top) and FIRE-2 (bottom) simulations.
Galaxy ID, stellar mass and SFR are noted in each panel.
The dashed white circle ($R = 12.4$\,kpc) indicates the median impact parameter of CGM-reddened galaxies from JADES observations. 
The legend in the bottom-right corner shows the dust surface densities ($\Sigma_\mathrm{dust}$) measured from our observations, with the same color-coding as presented by the color bar in the top-right corner.
}
\label{fig:zmap}
\end{figure*}

\subsection{Comparison with Hydrodynamical Simulations}
\label{ss:04d_hydro}

In hydrodynamical simulations, it is usually difficult to transport very small dust grains to large distances ($\gtrsim10$\,kpc) from the star-forming disks.
Most recently, using the Cholla \citep{schneider15} hydrodynamics code, \citet{richie25} studied the dust grain transport in the CGM through multi-phase outflows.
Because of sputtering, small grains (sizes $a \sim0.001$\,\micron) are found to be quickly destroyed in all but the coolest gas, and their simulations predict a low dust surface density $\Sigma_\mathrm{dust} \lesssim 10^{-3}$\,\msunsqpc\ including grains of all sizes.
This is significantly smaller than the dust surface density that we observed ($\Sigma_\mathrm{dust} \sim 10^{-1}$\,\msunsqpc) assuming a conversion factor of $\Sigma_\mathrm{dust} / A_V = 0.2$\,\si{M_\odot.pc^{-2}.mag^{-1}} \citep[the same as that assumed by ][]{mccleary25}.
We note that the galaxy simulated by \citet{richie25} is much lower in stellar mass ($M_\mathrm{star} = 10^{10}$\,\msun) than DSFGs that we study, and we caution that such a comparison could be unfair.

Most hydrodynamical simulations do not include live, self-consistent dust grain components themselves.
Therefore, we directly compare the simulated surface density profile of gas ($\Sigma_\mathrm{gas}$) and gas-phase metals ($\Sigma_\mathrm{Z,gas}$) with our observations of $\Sigma_\mathrm{dust}$ in the CGM.
As the dust-to-metal ratio (D/M) is strictly below 1, $\Sigma_\mathrm{Z,gas}$ is considered as an upper limit of $\Sigma_\mathrm{dust}$ from these simulations.
Figure~\ref{fig:zmap} shows the $\Sigma_\mathrm{Z,gas}$ maps of massive star-forming galaxies ($M_\mathrm{star} \sim 10^{11}$\,\msun\ and $\mathrm{SFR} > 200$\,\smpy) at the $z = 3.5$ snapshot in two hydrodynamical simulations, IllustrisTNG \citep[specifically, TNG100-1;][]{nelson18,pillepich18,springel18} and FIRE-2 \citep[specifically, the massive halo suite;][]{aa17, cochrane23, wetzel23,wetzel25}.
The $M_\mathrm{star}$ and SFR of these simulated galaxies are comparable to those of the foreground DSFGs within our sample \citep[e.g.,][]{dacunha15, gomez22, hodge25}.
In both simulation suites, the gas-phase metals in massive star-forming galaxies are clearly concentrated in the compact star-forming disks.
At $R = 12.4$\,kpc, the median impact parameter of CGM-reddened galaxies from our observations, the surface densities of gas-phase metals are typically low ($\Sigma_\mathrm{Z,gas} \simeq 10^{-2} - 10^{0}$\,\msunsqpc), exhibiting anisotropic 2D density profiles as a result of the complex merger and feedback history.

\begin{figure*}[!t]
\centering
\includegraphics[width=\linewidth]{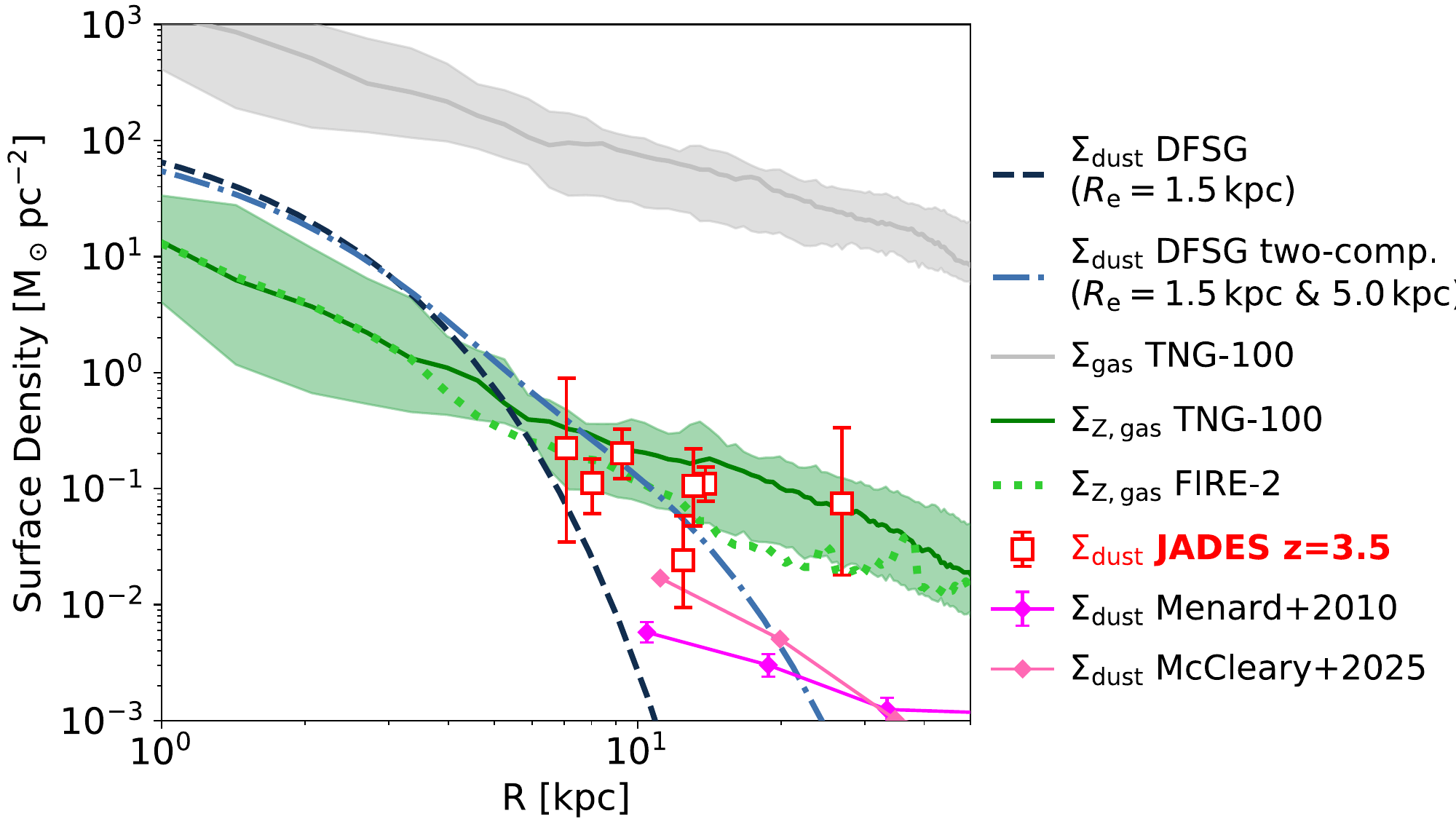}
\caption{Comparisons of surface density profiles of gas ($\Sigma_\mathrm{gas}$), gas-phase metals ($\Sigma_\mathrm{Z,gas}$) and dust ($\Sigma_\mathrm{dust}$) for DSFGs at cosmic noon.
$\Sigma_\mathrm{dust}$ measured through JADES $z=3.5$ CGM sightlines are highlighted as open red squares.
$\Sigma_\mathrm{dust}$ profiles inferred from CGM dust extinction at $z \lesssim 0.3$ (\citealt{menard10}, magenta; \citealt{mccleary25}, pink) are shown for comparison.
We also plot the radial profiles of $\Sigma_\mathrm{gas}$ (silver) and $\Sigma_\mathrm{Z,gas}$ (green) from the massive star-forming galaxies at $z=3.5$ in TNG100-1 (solid lines) and the FIRE-2 massive halo suite (dotted line) simulations (Figure~\ref{fig:zmap}).
Finally, $\Sigma_\mathrm{dust}$ profiles of massive DSFGs ($M_\mathrm{dust} = 10^9$\,\msun) assuming single exponential component ($R_\mathrm{e} = 1.5$\,kpc; dashed dark-blue line) and two exponential components ($R_\mathrm{e} = 1.5$\,kpc for 80\% of the mass, and $R_\mathrm{e} = 5.0$\,kpc for the remaining 20\%; dash-dotted blue line) are shown for comparison.
}
\label{fig:dust_prof}
\end{figure*}

The radial profiles of $\Sigma_\mathrm{Z,gas}$ in both simulations are extracted and displayed in Figure~\ref{fig:dust_prof}.
The 0-50-100th percentiles of the $\Sigma_\mathrm{Z,gas}$ profiles are extracted from all galaxies in the $z=3.5$ snapshot that host SFR greater than 200\,\smpy, including seven galaxies in TNG100-1 and four galaxies in the FIRE-2 massive halo suite.
Surprisingly, we find that our $\Sigma_\mathrm{dust}$ measurements are comparable to those of $\Sigma_\mathrm{Z,gas}$ in simulations at the same radius.
However, the D/M in the CGM should be much lower than unity because of destruction processes such as sputtering.
For example, the dust evolution model by \citet{otsuki24} suggests $\mathrm{D/M} \lesssim 0.1$ within a timescale of 1--2\,Gyr. 
If correct, this would imply that the observed $\Sigma_\mathrm{dust}$ is $\sim10\times$ higher than the predictions from simulations.

The prominent mismatch between observations and simulations is likely caused by the underestimated D/M and/or gas-phase metallicity in the CGM.
At $R \sim 10$\,kpc, the gas-phase metallicity in TNG100-1 massive star-forming galaxies is typically sub-solar ($Z \sim0.2 Z_\mathrm{\odot}$).
We argue that if the gas at this distance is enriched to solar metallicity, for example if the gas is dominated by the outflow launched from an enriched starburst cloud, we may reproduce the observed $\Sigma_\mathrm{dust}$ with a D/M also similar to those observed or simulated in the ISM of massive galaxies ($\mathrm{D/M} \simeq 0.2 - 0.5$; e.g., \citealt{devis19,liqi19,konstantopoulou24,algera26}).

\subsection{Comparison with ALMA Observations}
\label{ss:04d_alma}

So far, the majority of ALMA (sub-)millimeter observations of DSFGs at cosmic noon suggest compact sizes of dust continuum emission, typically with small half-light radii of $R_\mathrm{e} =$\,1--2\,kpc \citep[e.g.,][]{simpson15, hodge16, hodge19, elbaz18, fujimoto18, gullberg19} and high surface brightness (surface IR luminosity $\Sigma_\mathrm{IR} \gtrsim 10^{12}$\,\si{L_\odot.kpc^{-2}}).
However, extended dust continuum emission out to $R \sim 5$\,kpc is also observed for certain DSFGs on the so-called ``star-forming main sequence'' \citep[e.g.,][]{chengc20,tadaki20,sunf21a, gomez22b} with relatively low surface luminosity ($\Sigma_\mathrm{IR} \lesssim 10^{10}$\,\si{L_\odot.kpc^{-2}}).
In particular, \citet{gullberg19} conducted a stacking analysis of ALMA dust continuum observations of DSFGs in the visibility plane, finding an additional extended dust continuum on $\sim 4$\,kpc scales which contributes $\sim13\%$ of the total 870-\micron\ emission.
With low surface brightness ($\lesssim 0.1$\,\si{mJy.arcsec^{-2}} in ALMA Band 7), such an extended continuum emission can elude detection by ALMA blank-field surveys especially with extended configurations (i.e., small synthesized beam size).

In Figure~\ref{fig:dust_prof}, we also compare the radial $\Sigma_\mathrm{dust}$ profile measured through CGM extinction with that inferred from ALMA observations.
Motivated by aforementioned literature especially \citet{gullberg19}, we assume two profiles with fixed total dust mass $M_\mathrm{dust} = 10^9$\,\msun.
Such a dust mass is consistent with those of the foreground DSFGs in our sample \citep[e.g.,][]{dacunha15, hodge25}.
In a single exponential component profile with $R_\mathrm{e} = 1.5$\,kpc, $\Sigma_\mathrm{dust}$ will rapidly drop below $10^{-2}$\,\msunsqpc\ at $R > 10$\,kpc, and thus not match our observations.
However, if we assume that 20\%\ of the dust mass is distributed in a secondary extended exponential profile with $R_\mathrm{e} = 5.0$\,kpc, we would expect $\Sigma_\mathrm{dust} \sim 10^{-1}$\,\msunsqpc\ at the median impact parameter of CGM-reddened galaxies in our sample, and thus match our observations.

We note that the inferred $\Sigma_\mathrm{dust}$ along the sightline of ID 81034 at $R\sim27$\,kpc still remains as an outlier to the DSFG two-component dust model.
However, we caution that the expected low surface brightness ($\Sigma_\mathrm{870\,\mu m} \sim 10^{-2}$\,\si{mJy.arcsec^{-2}}) at $R = $\,3\farcs7 is below the detection limit and beyond the recoverable scale of the Band-7 stacking experiment by \citet{gullberg19}, and thus remains virtually unexplored.
As suggested by the ALMA online sensitivity calculator, detecting 870-\micron\ continuum emission at such a low surface brightness over a synthesized beam size of 1\farcs0 (i.e., achievable with the most compact ALMA 12\,m array with the best sensitivity to extended emission) at S/N\,$\geq4$ per beam would require on-source integration time over $\sim 100$\,hours.
Therefore, we conclude that our observations suggest the presence of extended dust reservoirs in the CGM of DSFGs at $z\sim3.5$, which are largely invisible to contemporary (sub-)millimeter interferometers such as ALMA because of the low surface brightness ($\Sigma_\mathrm{870\,\mu m} \sim 10^{-2}$\,\si{mJy.arcsec^{-2}}).

\subsection{Implication for the Physics of Dust Transport}
\label{ss:04e_phys}

The discovery of large reservoirs of small dust grains (and likely silicates) in the CGM of massive galaxies at $z = 3.5$ presents challenges to our current understanding of dust at high redshifts.
We highlight the following contrasts to known astrophysics of dust and DSFGs:
\begin{itemize}
    \item The extent of dust mass in DSFGs is found to be compact (effective radii $R_\mathrm{e} = 1-2$\,kpc), while we observe dust extinction in the CGM out to $R \sim 30$\,kpc;
    
    \item The dust attenuation curves ($A_\lambda / A_V$) of high-redshift galaxies as observed by JWST \citep[e.g.,][]{markov25a, markov25, mcKinney25, shivaei25} are typically found to be shallower than the SMC's, and similar to or even shallower than that of the \citet{calzetti00} attenuation curve, while the observed CGM dust extinction curve is similar to or steeper than that of the SMC. Note that we caution the difference between attenuation and extinction; also, the simple slab geometry for CGM dust extinction is different from complex stellar--ISM geometry for attenuation curve inference.
    
    \item The small dust grains inferred from the steep extinction would find it difficult to survive in a hot-phase outflow launched into CGM because of sputtering \citep[e.g.,][]{richie25}.
\end{itemize}


To reproduce the observed large reservoirs of small dust grains in the CGM, multiple physical processes are necessary to (1) ensure the survival of small dust grains through the transport from ISM to CGM; (2) effectively destroy large and carbonaceous dust grains, or reduce their transport; and (3) efficiently transport dust grains to $\sim$30\,kpc within a timescale of $\sim 1$\,Gyr (i.e., the lookback time from $z=3.5$ to $z = 7$).

Below we discuss a few physical processes that might be responsible to reproduce the observed dusty CGM:
\begin{enumerate}
\item \textit{Galaxy merger and interactions}. 
We stress that all of our observational evidence of large dust reservoirs in the CGM is associated with five DSFGs in two galaxy protoclusters at $z=3.47$ and 3.70.
Galaxies in protoclusters feature an enhanced rate of mergers and interactions because of high number density and relatively low velocity dispersions \citep[e.g.,][]{hine16, lius25}.
Galaxies during and post merger feature enhanced gas and metal surface density in the CGM \citep[e.g.,][]{hani18, ginolfi20, dicesare24} through merger-induced outflows and tidal strippings, thus enhancing the surface density of dust.

\item \textit{Cool gas outflows}. Among the multi-phase gas outflows that enrich the CGM, the cool-phase ($T \lesssim 10^4$\,K) outflows could be particularly important for CGM dust enrichment.
The survival of dust grains in hot winds is challenging, but cooler dense clumps entrained in outflows are much more favorable for dust survival through mechanisms like self-shielding and radiative cooling \citep[e.g.,][]{farber22, chenz24}.
The recent JWST discovery of warm dust in the CGM of the \textit{Makani} galaxy ($z=0.459$) out to $R\sim35$\,kpc, inferred to have been launched by a starburst wind, suggests that cool outflowing dust can survive the long transport into the CGM \citep{veilleux25}.

\item \textit{Dust shattering and sputtering}. 
The fragmentation of large grains through grain-grain collisions is likely the key to reproducing the observed steep extinction curve through the enhancement of small grains.
Turbulent cool clumps in the CGM might be responsible for efficient dust shattering and increasing the fraction of grains with sizes $a \lesssim 0.01$\,\micron\ \citep[e.g.,][]{hirashita21, hirashita24}.
On the other hand, small dust grains are especially vulnerable to sputtering in the hot ($T \gtrsim 10^6$\,K) diffuse CGM, and therefore it is necessary to balance among sputtering, shattering and new dust supply from outflows to match the observed steep dust extinction.

\item \textit{Radiation pressure}. 
Radiation pressure from stellar light provides a physically plausible mechanism for dust transport into the CGM.
While gravitational force on a dust grain scales as $a^{3}$ (through its mass, where $a$ is the grain size), radiation pressure scales as $a^{2}$ (through the geometric cross section), suggesting that smaller grains are, in principle, more efficiently accelerated. 
In the central regions of local ultraluminous infrared galaxies and high-redshift DSFGs, the radiation field can approach the Eddington limit set by radiation pressure on dust \citep[e.g.,][]{thompson05, barcos-munoz17, hodge19}.
The radiation-driven dust transport can be grain-size selective. 
\citet{hirashita19} show that intermediate-sized grains ($a \sim 0.1$\,\micron) are most favored because smaller grains are more likely trapped in the disk because of gas drag.
Subsequent processing within the CGM such as shattering and sputtering will be required to reshape the grain size distribution and thus reproduce the observed extinction curve.

\end{enumerate}

\begin{figure}[!t]
\centering
\includegraphics[width=\linewidth]{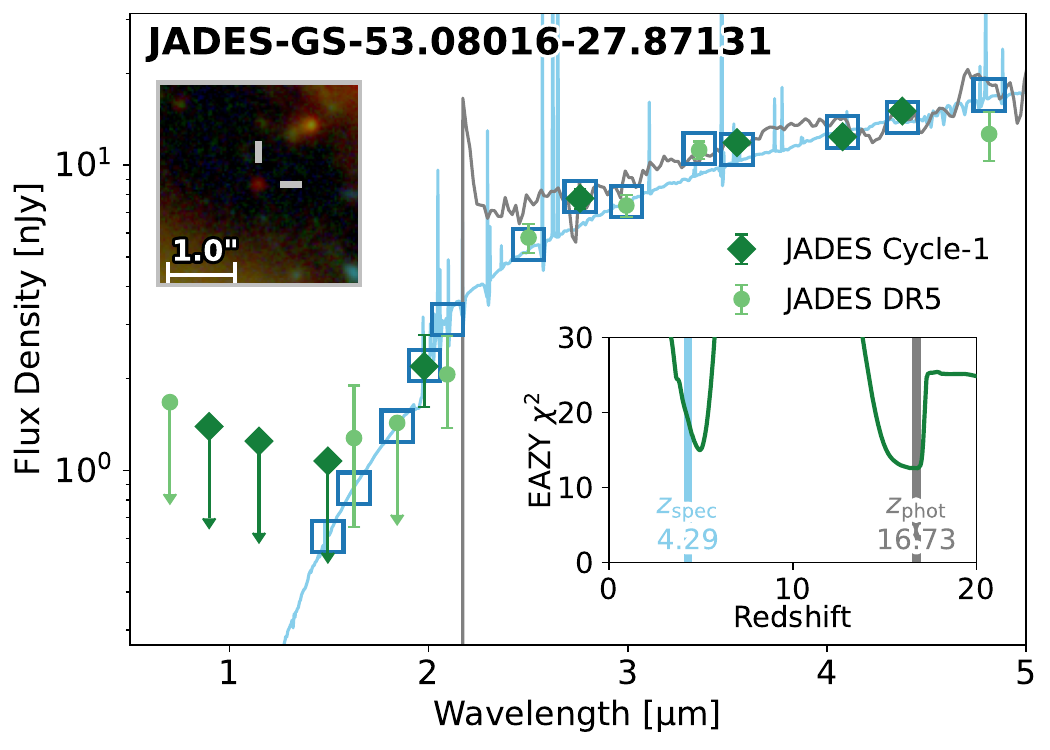}
\caption{A CGM-reddened galaxy at $z=4.29$ whose SED resembles that of a Lyman-break galaxy at $z>16$ \citep{hainline24_z8, deugenio25d}.
Inset panels show the NIRCam F444W--F200W--F115W RGB image and $\chi^2(z)$ curve from EAZY SED modeling \citep{eazy} using Cycle-1 photometry \citep{eisenstein25_jof}.
JADES Cycle-1 and Cycle-2/3 photometric measurements are shown as dark-green diamonds and light-green circles, respectively (upper limits are $2\sigma$).
The best-fit SED with Cycle-1 photometry indicates $z_\mathrm{phot} = 16.73$ (solid gray line; \citealt{hainline24_z8}).
The best-fit SED (fixed at $z=4.29$) using an emission-line galaxy template and power-law CGM dust extinction is shown as the skyblue line, with modeled photometry displayed as open blue squares.
}
\label{fig:z16}
\end{figure}

\subsection{Implication for the Surveys of Galaxies at Very High Redshifts ($\mathit{z \gtrsim 10}$)}
\label{ss:04f_z10}

Finally, we discuss the implication of dusty foreground CGM on the observational search and studies of galaxies at very high redshifts ($z \gtrsim 10$).
Because of the sharp rise of the CGM dust extinction curve at rest-frame UV wavelengths, background galaxies at $z\simeq 3 - 7$ can be heavily attenuated in these shorter wavelengths, exhibiting strong continuum breaks that may resemble the \lya\ break at $z \gtrsim 10$.

Such a scenario is particularly highlighted by JADES-GS-53.08016-27.87131 in Figure~\ref{fig:z16}.
This source is projected 3\farcs0 to the southeast of ALESS10.1 (Figure~\ref{fig:aless10_1}), i.e., the same direction as ID 171525 but with an even larger impact parameter.
With eight-band NIRCam photometry obtained in Cycle-1, this galaxy was selected as a Lyman-break galaxy candidate at $z_\mathrm{phot} = 16.74$ with clear F200W dropout \citep{hainline24_z8}.
However, the authors cautioned that the fidelity was low because of the reddened UV SED, unusual of galaxies at the redshift frontier which are typically low in dust content.
The collection of deep medium-band imaging data over Cycles-2/3 further favors the low-redshift solution at $z<5$, and therefore the source has been rejected from the sample of JADES high-redshift dropout galaxies \citep[e.g.,][]{robertson24, whitler25}.
Most recently, \citet{deugenio25d} obtained deep NIRSpec G235M and G395M spectroscopy of this source through the JADES ``Dark Horse'' pilot survey, confirming the redshift at $z=4.29$ through the robust detection of \ha\ emission.
Using the same method outlined in Section~\ref{ss:03d_ext}, we successfully reproduce the observed SED with high dust attenuation $\log(\tau_V) = 0.30\pm0.15$ with a SMC-like extinction curve ($n = -1.2 \pm 0.5$).
The large $\tau_V$, if true, would exceed those of all sources in our sample at such a large impact parameter of $R = 22$\,kpc.
However, we caution that this faint galaxy would not satisfy the NIRCam WFSS selection, and therefore the dust attenuation modeling using the spectral templates from brighter WFSS-selected galaxies could be misleading in this context.

In conclusion, we highlight that the foreground CGM dust extinction may redden galaxies in the background, such that their SEDs may satisfy the color selection for Lyman-break galaxies at $z \gtrsim 10$ (see further discussion by \citealt{eisenstein26}).
Caution should be taken for the interpretation of dropout galaxies with close angular separations ($\lesssim 4$\arcsec) from foreground massive galaxies, especially when the number of photometric bands is limited and the observed SEDs are red.

Finally, we also note that the luminous galaxy JADES-GS-z14-0 at $z=14.18$ \citep[][]{carniani24, carniani25, helton25_z14,schouws25} is found in the background of a galaxy at $z=3.475$ (projected separation 0\farcs4), which resides in the same galaxy protocluster studied throughout this work.
JADES-GS-z14-0 is also found to host a slightly reddened UV continuum (UV slope $\beta = -2.20 \pm 0.07$), which has been interpreted as the dust attenuation from its own ISM ($A_V = 0.31_{-0.07}^{+0.14}$; \citealt{carniani24}).
The presences of non-negligible $A_V$, damped \lya\ absorption and high escape fraction of ionizing photons ($f_\mathrm{esc}\sim 20$\%; for the interpretation of weak UV emission lines;  \citealt{carniani25}) may require complex geometry or covering factor, potentially related to its outflow history as suggested by \citet{ferrara25}.
Although the ALMA dust non-detection at rest-frame 88\,\micron\ \citep{carniani25, schouws25} is fully compatible with the inferred $A_V$ \citep[see also][]{ferrara25}, we hypothesize the potential, non-negligible $A_V$ contribution from the foreground galaxy.
Future deep ALMA observations of the dust continuum emission of JADES-GS-z14-0 will be the key to unravel the puzzling dust content and extreme luminosity of such a luminous galaxy at the cosmic redshift frontier.

\section{Summary}
\label{sec:05_sum}

In this work, we present a deep JWST/NIRCam imaging and grism spectroscopic survey of the JADES Origins Field through Cycle-3 Program 4540.
We report the grism spectroscopic redshifts of \fsun{1,445} emission-line galaxies across $z = 0 - 9$. 
The grism spectroscopic redshift catalog is publicly released through this work.
Our main findings are summarized as follows:

\begin{enumerate}

\item We confirm the presence of two prominent galaxy protoclusters at $z = 3.47$ and 3.70 in our survey footprint.
These two protoclusters contain 139 sources identified with our NIRCam grism spectroscopy, corresponding to 61$\pm$3\% of the emission-line galaxies that we identified across $3.4<z<3.8$.
Both filamentary structures are anchored by massive DSFGs that have been characterized through previous ALMA studies.

\item Five of these DSFGs have background emission-line galaxies in their vicinity that shown unusually reddened UV--optical continuum and strong \oiii and/or \ha\ emission lines at the same time.
We interpret this as heavy dust extinction from the CGM of foreground DSFGs at projected separations of 7--30 kpc.

\item Through modeling the dust extinction curve from the observed SED of CGM-reddened galaxies, we infer high $A_V$ up to $\sim1$ magnitude and therefore large reservoirs of dust grains with high surface densities of $\Sigma_\mathrm{dust} \gtrsim 10^{-1}$\,\msunsqpc.
The surface density substantially exceeds those measured in low-redshift halos by a factor of $\sim10$.

\item The modeled dust extinction curves are found to be comparable to or even steeper than that of the SMC.
We do not detect any prominent 2175-\AA\ bump in the extinction curves.
We infer that the dust in the CGM of these galaxies is dominated by small silicate dust grains and has a deficit of large carbonaceous grains.
This is likely a result of the dust production by high-mass AGB evolution and CCSNe through the early starburst phases of DSFGs. 

\item The observed high surface density of $\Sigma_\mathrm{dust}$ in the CGM is comparable to the surface density of gas-phase metals $\Sigma_\mathrm{Z,gas}$ of massive star-forming galaxies in hydrodynamical simulations such as TNG100-1 and FIRE-2.
To match the observed $\Sigma_\mathrm{dust}$ profile with reasonable dust-to-metal ratios, the CGM of DSFGs will need to be enriched to $\sim$ solar metallicity. 

\item The large dust reservoir in the CGM suggests that DSFGs host an extended dust component ($R_\mathrm{e} \sim 5$\,kpc; 10--20\%\ of the total dust mass) in addition to the luminous compact component ($R_\mathrm{e} = 1 - 2$\,kpc; 80--90\%\ of the total dust mass) that is frequently revealed by ALMA continuum imaging.
Such an extended dust component in the CGM is largely invisible to (sub-)millimeter interferometers because of their low surface brightness (i.e., $10^{-2} \sim 10^{-1}$\,\si{mJy.arcsec^{-2}} at 870\,\micron).

\item We discuss the key physical processes that might be responsible for transporting a large amount of small dust grains to the CGM in these DSFGs, including (1) galaxy mergers and interactions enhanced by the protocluster environment, (2) cool gas outflows for dust entrainment and survival, (3) efficient dust shattering but limited sputtering, and (4) the combination of radiation pressure for the transport of intermediate-sized grains and subsequent grain processing in CGM. 

\item Finally, we caution that foreground CGM dust extinction may redden galaxies in the background at intermediate redshifts through a steep extinction curve, producing a strong continuum break that may resemble that of Lyman-break galaxies at $z \gtrsim 10$.
Caution must be taken when interpreting dropout galaxies with close angular separation from foreground massive galaxies, especially when the number of photometric bands is limited and the observed SEDs are red.

\end{enumerate}


\section*{acknowledgments}

FS acknowledges fruitful discussions with Suoqing Ji, Mingyu Li, Hui Li, Helena M. Richie and Yunjing Wu.
FS, DJE, KH, JMH, BDJ, MR, BR, EE, ZJ, CNAW and YZ acknowledge the support from JWST/NIRCam contract to the University of Arizona, NAS5-02105.
DJE is supported as a Simons Investigator.
RM and FDE acknowledge support by the Science and Technology Facilities Council (STFC), by the ERC through Advanced Grant 695671 “QUENCH”, and by the UKRI Frontier Research grant RISEandFALL. 
ST acknowledges support by the Royal Society Research Grant G125142.
AJB and JC acknowledge funding from the ``FirstGalaxies" Advanced Grant from the European Research Council (ERC) under the European Union’s Horizon 2020 research and innovation programme (Grant agreement No. 789056)
ECL acknowledges support of an STFC Webb Fellowship (ST/W001438/1).
RM also acknowledges funding from a research professorship from the Royal Society. JAAT acknowledges support from the Simons Foundation and JWST program 3215. Support for program 3215 was provided by NASA through a grant from the Space Telescope Science Institute, which is operated by the Association of Universities for Research in Astronomy, Inc., under NASA contract NAS 5-03127.
The research of CCW is supported by NOIRLab, which is managed by the Association of Universities for Research in Astronomy (AURA) under a cooperative agreement with the National Science Foundation.
JW gratefully acknowledges support from the Cosmic Dawn Center through the DAWN Fellowship. The Cosmic Dawn Center (DAWN) is funded by the Danish National Research Foundation under grant No. 140.
Funding for this research was provided by the Johns Hopkins University, Institute for Data Intensive Engineering and Science (IDIES)
The authors acknowledge use of the lux supercomputer at UC Santa Cruz, funded by NSF MRI grant AST 1828315.

This work is based on observations made with the NASA/ESA/CSA James Webb Space Telescope. The data were obtained from the Mikulski Archive for Space Telescopes at the Space Telescope Science Institute, which is operated by the Association of Universities for Research in Astronomy, Inc., under NASA contract NAS 5-03127 for JWST. These observations are associated with program \#1180, 1210, 3215, 4540.
Support for program \#3215 was provided by NASA through a grant from the Space Telescope Science Institute, which is operated by the Association of Universities for Research in Astronomy, Inc., under NASA contract NAS 5-03127.

This paper makes use of the following ALMA data: ADS/JAO.ALMA\#2015.1.00242.S and \#2018.1.01079.S. ALMA is a partnership of ESO (representing its member states), NSF (USA) and NINS (Japan), together with NRC (Canada), NSTC and ASIAA (Taiwan), and KASI (Republic of Korea), in cooperation with the Republic of Chile. The Joint ALMA Observatory is operated by ESO, AUI/NRAO and NAOJ. The National Radio Astronomy Observatory is a facility of the National Science Foundation operated under cooperative agreement by Associated Universities, Inc.

The authors acknowledge the use of ChatGPT (OpenAI) for language polishing; all scientific content and conclusions were produced by the authors.

\begin{contribution}

FS led the project conceptualization, data analyses and writing of the paper.
DJE led the key designs of the 4540 program as PI and administered the project.
FDE contributed to the project conceptualization. 
BDJ, BR, ST, DJE, and KH led the NIRCam imaging data processing and cataloging, with contribution from all authors. 
JMH and XL contributed to the NIRCam grism data analyses.
MR led the NIRCam instrument science team as PI.
MR, CNAW, EE, FS, KH, CCW contributed to the design, construction and commissioning of NIRCam. 
All authors contributed to the interpretation of results and writing of the paper.

\end{contribution}

\facilities{JWST (NIRCam), HST (ACS and WFC3-IR), ALMA
}

\software{Astropy \citep{2013A&A...558A..33A,2018AJ....156..123A,2022ApJ...935..167A},  CIGALE \citep{noll09, cigale},
          }

\setcounter{figure}{0}
\renewcommand{\thefigure}{\thesection\arabic{figure}}

\appendix
\restartappendixnumbering

\section{NIRCam WFSS Spectra and Redshift Catalog}
\label{apd:01_z}

\begin{figure}[!ht]
\centering
\includegraphics[width=\linewidth]{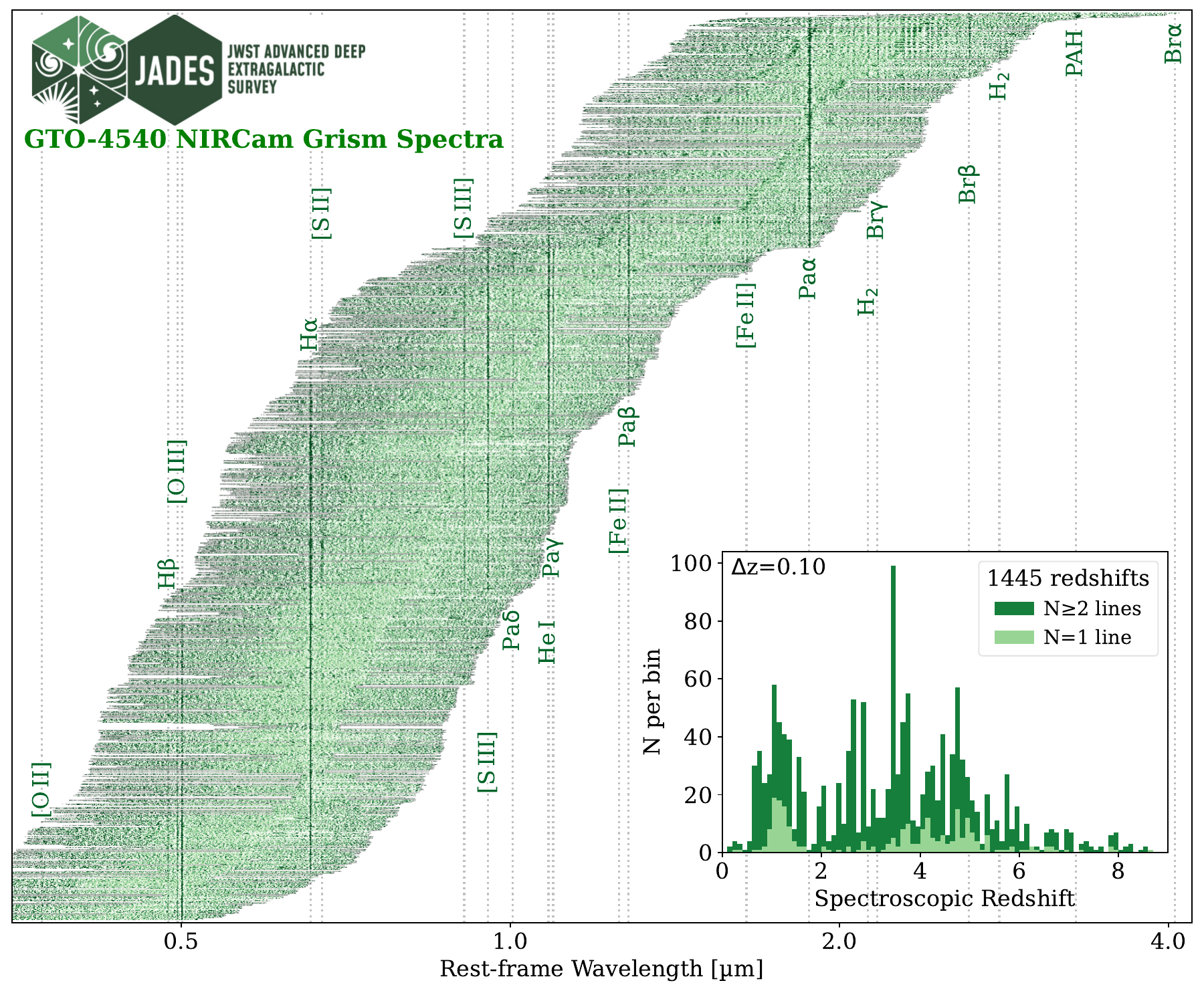}
\caption{JADES Cycle-3 GTO-4540 NIRCam WFSS emission-line spectra (F322W2+F356W+F444W combined) stacked and ordered by redshift. The prominent emission lines from \oii, \hb\ and \oiii\ at $z\sim8.7$ to \bra\ at $z \sim 0$ are indicated by vertical dotted lines. 
The inset panel shows the redshift histogram of confirmed emission line galaxies with a bin size $\Delta z= 0.10$.
}
\label{fig:spec-stack}
\end{figure}

\begin{figure}[!t]
\centering
\includegraphics[width=\linewidth]{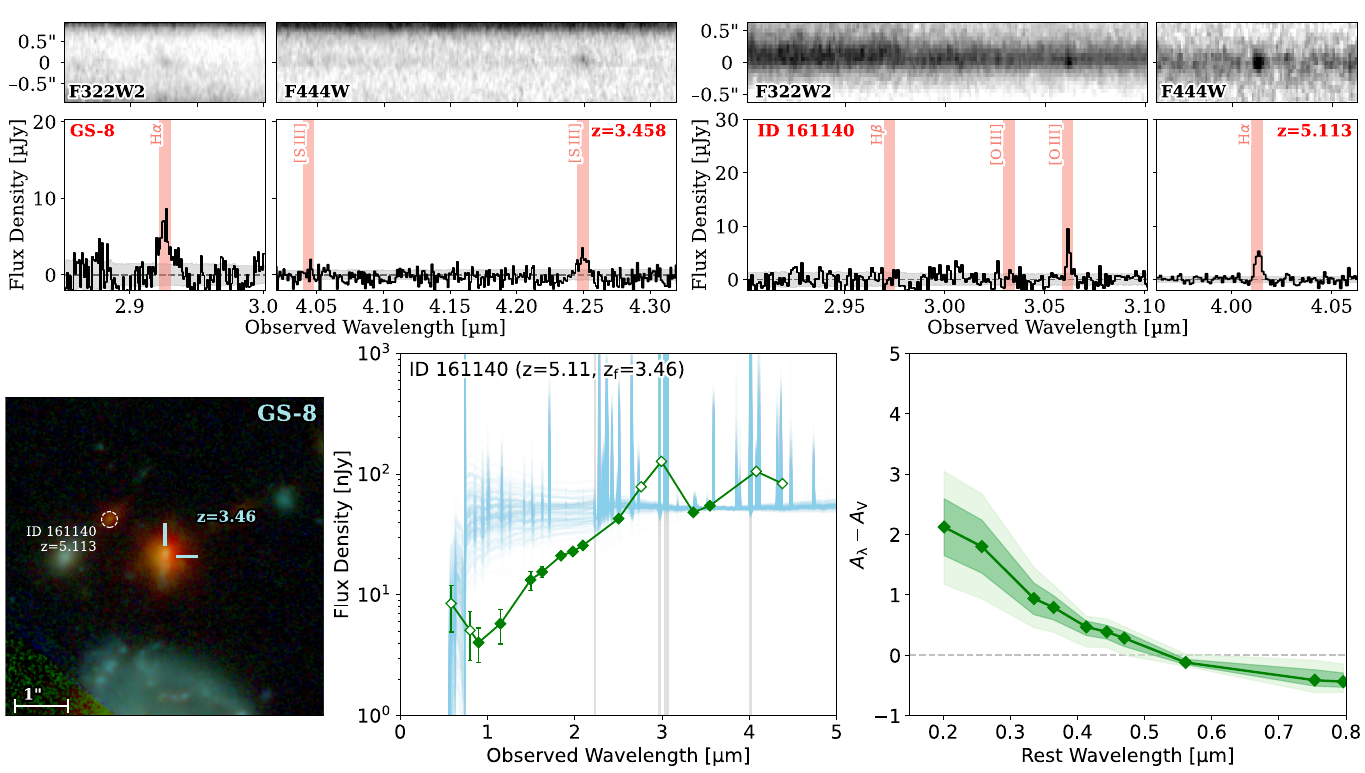}
\caption{Same as Figure~\ref{fig:aless10_1} but for JADES NIRCam ID 161140 in the background of GS-8 \citep[][NIRCam WFSS redshift $z=3.46$]{mckay25}.}
\label{fig:gs-8}
\end{figure}

\begin{figure}[!ht]
\centering
\includegraphics[width=\linewidth]{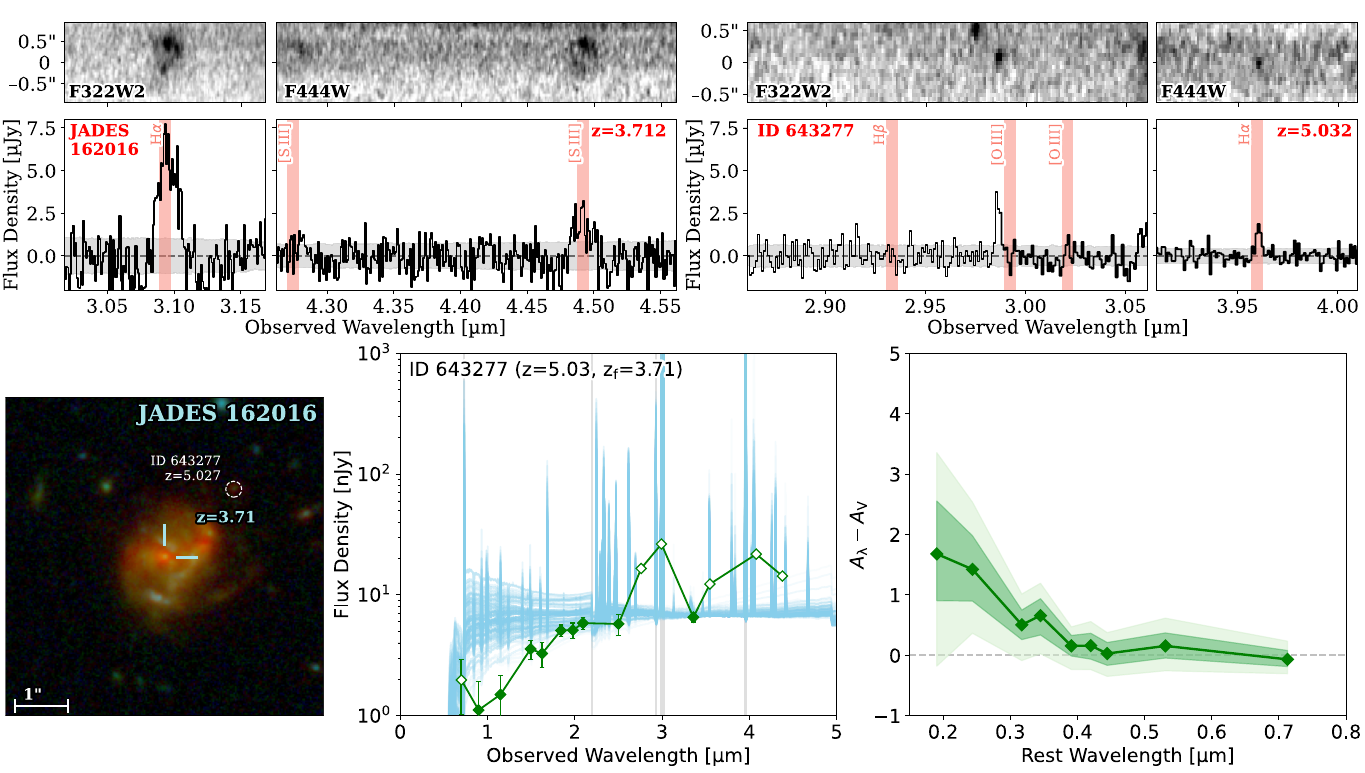}
\caption{Same as Figure~\ref{fig:aless10_1} but for JADES NIRCam ID 643277 in the background of JADES NIRCam ID 162016 (NIRCam WFSS redshift $z=3.71$).}
\label{fig:162016}
\end{figure}

\begin{figure}[!ht]
\centering
\includegraphics[width=\linewidth]{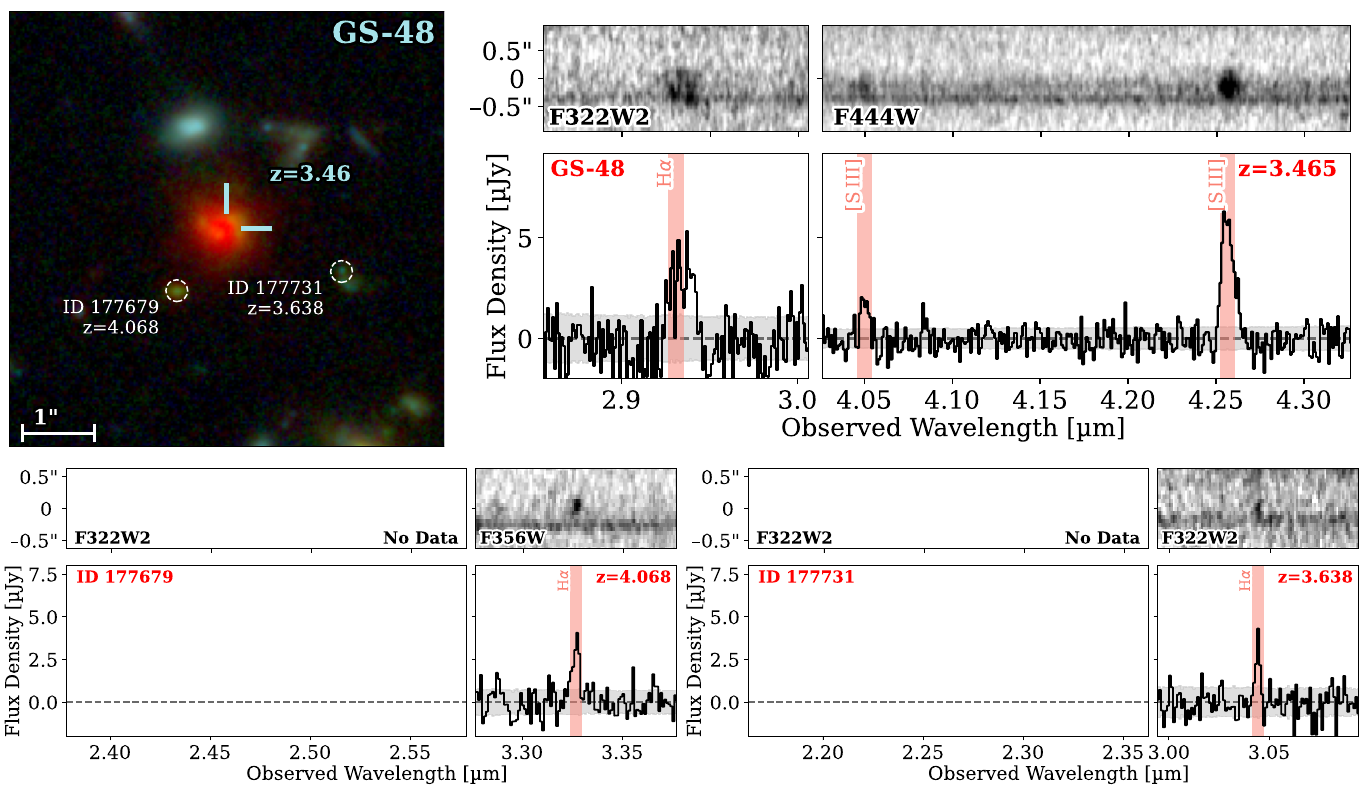}
\includegraphics[width=\linewidth]{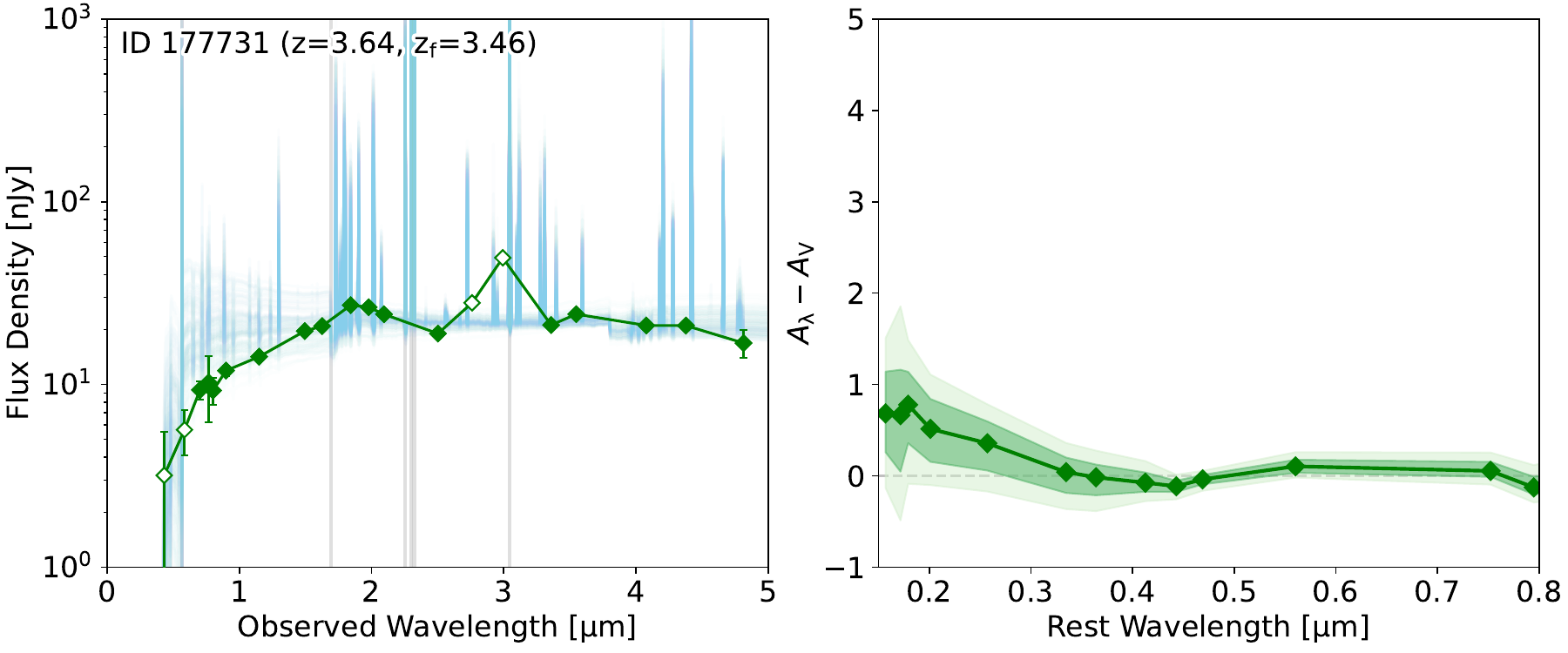}
\caption{Same as Figure~\ref{fig:aless10_1} but for JADES NIRCam ID 177679 and 177731 in the background of GS-48 \citep[NIRCam WFSS redshift $z=3.46$;][]{mckay25}.
Note that the observed SED and extinction curve of ID 177679 are shown in Figure~\ref{fig:ext_curve}.
}
\label{fig:gs-48}
\end{figure}

Figure~\ref{fig:spec-stack} shows the stack of JADES Cycle-3 GTO-4540 NIRCam grism spectra of \fsun{1,445} sources with confirmed redshifts across $z=0 - 8.7$.
For each source, we combine their NIRCam 1D spectra taken with the F322W2, F356W and F444W block filters.
Prominent strong emission lines from \oii, \hb, \oiii\ in the rest-frame optical to PAH 3.3\,\micron\ and \bra\ in the mid-IR are highlighted. 
\fsun{The NIRCam WFSS redshift catalog from GTO-4540 is made publicly available through a Zenodo data repository \href{https://doi.org/10.5281/zenodo.18328404}{https://doi.org/10.5281/zenodo.18328404}.}
Columns of this redshift catalog include source ID (\verb|ID|, matched to JADES DR5; \citealt{johnson26}, \citealt{robertson26}), coordinates (\verb|RA| and \verb|DEC|; unit: degree), spectroscopic redshifts (\verb|zspec|), confidence of \verb|zspec| (\verb|zconf|, 1--6; see Section~\ref{ss:03a_z} and \citealt{linx25a, sunf25b}), numbers of emission lines detected at $\mathrm{S/N} \geq 3$ (\verb|nlines|), names of detected emission lines (\verb|name|), and observed wavelengths of detected emission lines (\verb|wave|; unit: \micron).

Figure~\ref{fig:gs-8}, \ref{fig:162016} and \ref{fig:gs-48} show the NIRCam RGB image, 2D and 1D spectra and the observed SEDs of the remaining three CGM-reddened galaxies (together with their foreground DSFGs) as part of our sample.


\bibliography{00_main}{}
\bibliographystyle{aasjournalv7}

\end{document}